\documentclass[12pt, draftclsnofoot, onecolumn]{IEEEtran}

\usepackage{graphicx}
\usepackage{latexsym}
\usepackage{amsfonts}
\usepackage{amssymb}
\usepackage{amsbsy}
\usepackage{amsmath}
\usepackage{multicol}
\usepackage{float}
\usepackage{subfigure}
\usepackage{algorithm}
\usepackage[dvips]{epsfig}

\usepackage{fancyhdr}
\usepackage{epsfig}
\usepackage{threeparttable}
\usepackage{epsf,epsfig}
\usepackage{amsthm}
\usepackage[noadjust]{cite}
\usepackage{dsfont}
\usepackage{color}
\usepackage{enumerate}
\usepackage{setspace}
\usepackage{bbm}

\usepackage{array}
\makeatletter
\newcommand{\thickhline}{%
  \noalign {\ifnum 0=`}\fi \hrule height 1pt
  \futurelet \reserved@a \@xhline
}
\newcolumntype{"}{@{\hskip\tabcolsep\vrule width 1pt\hskip\tabcolsep}}
\makeatother

\usepackage{amsmath}
\usepackage{algorithm}
\usepackage[noend]{algpseudocode}
\algrenewcommand{\algorithmiccomment}[1]{\hskip3em$\triangleright$ #1}

\makeatletter
\def\BState{\State\hskip-\ALG@thistlm}
\makeatother

\newtheorem{remark}{Remark}
\newtheorem{lemma}{Lemma}
\newtheorem{proposition}{Proposition}
\newtheorem{assumption}{Assumption}
\newtheorem{corollary}{Corollary}

\def\E{\mathsf{E}}

\def\SIR{\mathsf{SIR}}
\def\L{d_\ell}
\def\W{d_w}

\def\l{\left}
\def\r{\right}
\def\({\left(}
\def\){\right)}
\def\[{\left[}
\def\]{\right]}

\def\nn{\nonumber}

\usepackage{dblfloatfix} 

\title{\LARGE Sense-and-Predict: Harnessing Spatial Interference Correlation for Cognitive Radio Networks}

\begin{document}
\vspace {-15px}
\author{\vspace {-8px}Seunghwan Kim,
Han~Cha,
Jeemin~Kim,
Seung-Woo~Ko,
and~Seong-Lyun~Kim
\thanks{S. Kim, H. Cha, J. Kim, and S.-L. Kim are with the School of Electrical and Electronic Engineering, Yonsei University, Seoul, Korea (email: \{shkim, chan, jmkim\}@ramo.yonsei.ac.kr, slkim@yonsei.ac.kr), S.-W. Ko is with the Department of Electrical \& Electronic Engineering, The University of Hong Kong, Hong Kong, (email: swko@eee.hku.hk).}
\thanks{
Part of this work was presented at IEEE DySPAN 2017 \cite{dyspan}.}
}

\maketitle

\vspace {-50px}

\begin{abstract}
\vspace {-10px}
\emph{Cognitive radio} (CR) is a key enabler realizing future networks to achieve higher spectral efficiency by allowing spectrum sharing between different wireless networks. 
It is important to explore whether spectrum access opportunities are available, but conventional CR based on \emph{transmitter} (TX) sensing cannot be used to this end because the paired \emph{receiver} (RX) may experience different levels of interference, according to the extent of their separation, blockages, and beam directions. 
To address this problem, this paper proposes a novel form of \emph{medium access control} (MAC) termed \emph{sense-and-predict} (SaP), whereby each secondary TX predicts the interference level at the RX based on the sensed interference at the TX; this can be quantified in terms of a spatial interference correlation between the two locations.
Using stochastic geometry, the spatial interference correlation can be expressed in the form of a conditional coverage probability, such that \emph{the signal-to-interference ratio} ($\SIR$) at the RX is no less than a predetermined threshold given the sensed interference at the TX, defined as an \emph{opportunistic probability} (OP). The secondary TX randomly accesses the spectrum depending on OP. 
We optimize the SaP framework to maximize the \emph{area spectral efficiencies} (ASEs) of secondary networks while guaranteeing the service
quality of the primary networks. Testbed experiments using USRP and MATLAB simulations show that SaP affords higher ASEs compared with CR without prediction.\end{abstract}


\vspace {-20px}
\IEEEpeerreviewmaketitle

\section{Introduction}
\vspace {-8px}
Spectrum has become the most important and scarce resource during the period in which wireless communications have expanded enormously.
Many efforts have been made to extend the range of the usable spectrum even when it is already used by others; the research field is termed \emph{cognitive radio} (CR) \cite{HaykinCR}. 
By allowing spectrum-sharing among different wireless networks, CR is expected to increase the spectrum utilization efficiency several-fold. 
Furthermore, recent advances in spectrum sensing \cite{TutorialSensingCR} and multi-antenna techniques \cite{TutorialMIMOCR} indicate that CR is useful to facilitate massive connectivity, a vision of 5G communications \cite{5GMagazine}.

\begin{figure}
\centering 
{\includegraphics[width=10cm]{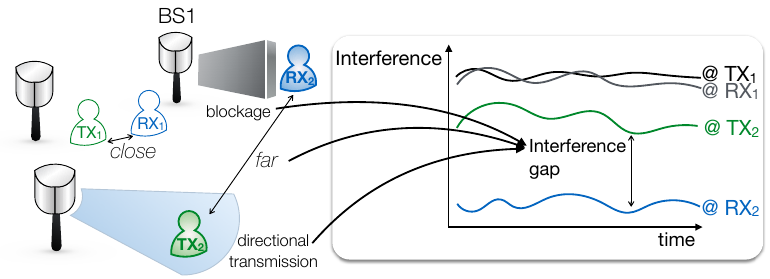}} \vskip-10pt
\caption{An illustration of the access opportunity detection problem because of: (i) secondary TX-RX distance separation; (ii) the blockages; and (iii) the directional signal transmissions. When secondary TX-RX association distance is long or when there are many blockages or primary network uses sophisticated beamforming, the $\SIR$ measured at TX significantly differs from the $\SIR$ at RX (see the $\SIR$ gap between $\textsf{TX}_2$ and $\textsf{RX}_2$), leading to detection errors.\label{system}}
\vspace{-.3in}
\end{figure}

CR features two types of users, primary and secondary, with the user status depending on their priorities within a given spectrum. 
A secondary \emph{transmitter} (TX) senses the medium prior to access to check the spectrum utilization by primary users. 
However, such a sensing mechanism (triggered by the TX) has the fundamental drawback that the interference level at the secondary TX differs from that at the \emph{receiver} (RX).
This interference gap may mislead the access decision of the secondary TX, causing transmission failure. 
To address this issue, we develop a novel \emph{medium access control} (MAC) termed {\it Sense-and-predict} (SaP), whereby each secondary TX decides to access the spectrum by predicting the interference level at its RX based on the sensed interference level at the TX. 
Specifically, the interference level at the RX is quantified as an \emph{opportunistic probability} (OP) defined as a conditional distribution of the interference level at the secondary RX given the measured interference at the TX; the probability that the secondary TX randomly accesses the medium is a function of the OP. 
Note that the OP is closely related to the spatial interference correlation between two separated locations, attributable to (i) the distance between the TX and the RX; (ii) blockages; and (iii) directional signal transmissions (see Fig.~\ref{system}).
We develop the spatial correlation in the form of a probability using \emph{stochastic geometry} (SG), and use our result to obtain an accurate OP and optimize the SaP framework.

\vspace {-10px}
 \subsection{Prior Works}
\vspace {-5px}

In the area of CR MAC, one research thrust focuses on integrating the techniques of sensing and communication into one framework, enabling to optimize the performance of secondary networks while guaranteeing the primary networks' requirement \cite{cmacSurvey}. It is further divided into different white space paradigms, e.g., interweave and underlay \cite{whitespace}. In the interweave approach, the secondary network attempts to access the medium only when the primary network is being inactive, called a spectrum hole, based on sensing and predicting the activities of the primary users \cite{surveyOCCUPANCY,surveyOCCUPANCY_2,surveyACTIVITYMODEL,SPECTRUM_PRIDICTION}. It is challenging to make the accurate prediction of the primary networks due to the lack of relevant information especially under the complex spectrum usage pattern in 5G, making it demanding to use the interweave approach. On the other hand, underlay approach has been recognized as a viable approach because the secondary users enable to access the medium even when the primary networks are active given the constraint of allowable level of interference. It leads to extensive studies on the underlay paradigm using a wide range of advanced techniques (e.g., power and admission control \cite{underlay1}, energy harvesting \cite{underlay2}, and, full-duplex \cite{underlay3}). The above works have considered the interference gap between a secondary TX and its paired RX to be negligible, which become rather invalid in 5G supporting stricter requirements than before. To guarantee the performance of secondary users, the interference gap between the secondary TX and the RX must be addressed.


Specifically, two types of problems are recognized: hidden and exposed node problems. 
A hidden node problem arises when an interferer is sensed not by the TX, but rather by the RX, and an exposed node problem is the reverse problem \cite{csma}. 
These are challenging issues in the field of \emph{carrier sensing multiple access} (CSMA) because they cause transmission failure due to a collision between two transmissions. To tackle these problems, most prior works have sought to implement additional control signals \cite{rts_cts,busytone_1,busytone_2}. 
In \cite{rts_cts}, a \emph{collision avoidance} (CA) scheme based on handshaking featured the transmission of control packets such as request-to-send, clear-to-send, and data-sending signals. In \cite{busytone_1}, a CA scheme based on a busy tone mechanism was proposed; the TX or RX sends a control packet to reserve the channel for transmission. This was extended in \cite{busytone_2} by using an asymmetrical dual busy tone mechanism to solve the exposed node problem. 
In CR, similar phenomenons arise depending on the locations of the primary TXs. To address this issue, for example, the authors in \cite{POMDP} designed different signaling protocols depending on whether the secondary TX and RX experience the same primary TXs or not. However, the approach of developing a new signal protocol can intensify control-plane congestion especially when supporting massive connectivity.
We use simple ALOHA-type random access, thus not increasing the signaling overhead to cope with the hidden/exposed node problems, but fine-tuning the access probability depending on the sensed interference at the TX; we use SG to this end. 

SG is accepted to be an efficient tool affording a tractable approach to interference modeling in large-scale wireless networks by focusing on the analysis of a single typical point \cite{Haenggi13}.
However, quantifying the spatial interference correlation using SG is far more challenging, because joint analysis of two typical points is required \cite{Ganti09, Shankar16,hskim}. 
The extent of correlation is highly dependent on the topological differences between the two points. 
For example, when the two nodes are co-located, it is obvious that they experience the same level of interference. 
As the distance between the nodes increases, the similarity decreases, and they finally become independent. 
The authors in \cite{Ganti09} described spatial correlation in the form of a correlation coefficient in ad hoc networks. 
Extending such work, \cite{Shankar16} focused on the joint coverage probability at two locations in a cellular network in which mobile devices were moving, and \cite{hskim} investigated the conditional interference distribution between two locations. 
However, no study has yet considered blockage and directional transmission effects on spatial interference correlations. 
These are more complicated because hidden/exposed node problems must be considered together. 

Blockages and directional signal transmissions render the \emph{line-of-sight} (LOS) conditions different at the two spatial locations, decreasing the mutual spatial interference correlation. 
In CR networks, this causes errors in OP detection. 
For example, when an interferer is in LOS to the secondary TX but not to the RX, an exposed node problem can occur. 
It makes the secondary TX lose the chance to access the medium although the real interference is small. 
In the opposite case, a hidden node problem occurs, causing transmission failure because interference is higher than expected. 
Both problems become more critical when \emph{millimeter-wave} (mmW) frequency is employed; signals are directionally transmitted with high power and now more vulnerable to blockage \cite{JH15}. 
The blockage effect in wireless networks has recently been investigated using SG \cite{Bai14,andrews_diversity,jm_access,kaifeng}.
Blockages were modeled using a Boolean approach \cite{Bai14,jm_access} or a random lattice model \cite{kaifeng}. 
However, such models consider the blockage effect only in terms of single-link connectivity.
Authors in \cite{andrews_diversity} focused on the blockage effects on multiple links but did not consider the effects thereof on spatial interference correlations among the links. 
Unlike the aforementioned works, our current study incorporates mmW effects when analyzing spatial correlations, allowing application of the SaP framework to mmW bands.

\vspace {-20px}
 \subsection{Contributions}
\vspace {-5px}
In this work, we tackle the interference mismatch problem in sensing-based underlay CR networks by exploiting the spatial correlation between the secondary TX and the RX, which is jointly affected by the locations of interferers, blockages and beamwidth of the networks. 
For example, as the secondary-pair distance increases, blockages and directional transmissions are likely to make more complex LOS conditions of interferers between them.
In a similar vein, as the number of blockages increases and/or the beamwidth decreases, the interference gap also increases though the pair distance is identical. 

Specifically, we consider a spatial sensing where a secondary TX’s sensing result depends on the geometry of the primary TXs~\cite{d2d}. Contrary to conventional temporal sensing observing primary users’ temporal behaviors for a relatively long duration~\cite{kwsung}, the spatial sensing relies on one sensing result assuming that primary TXs’ behaviors are unchanged during a concerned duration. It leads to an immediate access control of each secondary TX, which is one of the key requirements of 5G. On the other hand, the interference mismatch between TX and RX causes significant errors such as hidden/exposed node problems. It calls for developing an interference prediction model at the paired RX using a powerful tool of SG.
To the best of our knowledge, it is the first work considering the spatial correlation in cognitive radio MAC.

Our SG-based model captures such spatial correlations, thereby providing an access probability of the secondary TX. 
Specifically, the spatial correlation is provided in the form of a probability defined as the OP, and the secondary TX uses the OP to determine the access probability.
It is worth noting that this probabilistic approach enables a transformation of the real interference value to a value between $0$ to $1$, facilitating a random access design.
Furthermore, the approach provides the relative levels of interference caused by various parameters, such as TX density and transmission power. 
For example, even if the measured interference values are equal at two different places, the OPs are not identical because of the differences in the network parameters. OP enables a multiplicity of parameters to be accommodated within one metric in a probabilistic form, thus harmonizing the SaP framework. Our contributions are listed below.

 \begin{figure*}  
 	\vspace {-5px}
\centering
 \subfigure[Without $\SIR$ Prediction]{\centering
   \includegraphics[width=6.8cm]{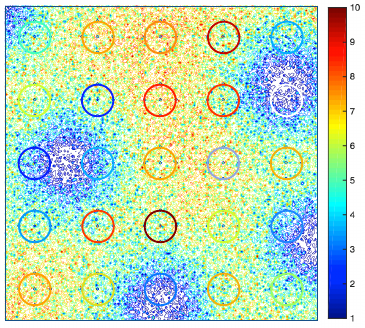} } 
   \quad
  \subfigure[With $\SIR$ Prediction]{
   \includegraphics[width=6.8cm]{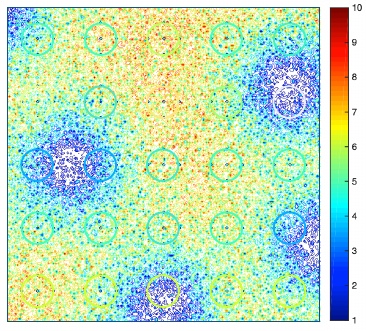}} \vskip-10pt
\caption{The $\SIR$ prediction where the densities of the primary TXs and blockages are 
 $10^3\text{ TXs}/\text{km}^2$ and $10^3\text{ buildings}/\text{km}^2$, respectively. The color represents $\SIR$ level at the corresponding location. Each secondary TX is located in the center of the circle. 
The circumference refers to the possible RX location separated by $2$ m,
of which the color is a) equal to its center in the case without $\SIR$ prediction and b) predicted $\SIR$, shown that the SIR prediction is so accurate that the circle is almost invisible. \label{Fig:SIRcompare}}\vspace{-.4in}
\end{figure*} 

\begin{itemize}
\item {\bf Spatial correlation analysis}: As mentioned, we capture the spatial interference correlation in the form of an OP, enabling to cope with different topologies of primary interferers within our probabilistic framework. It is different from other works designing a signaling protocol to address each case separately (e.g.,~\cite{POMDP}). Regarding the distinct characteristics of wireless signal propagation, we consider two separate scenarios: below- and above-$6$ GHz networks (e.g.,  mmW network). Specifically, for a below-$6$ GHz network, we consider only interferer locations when analyzing spatial correlations. For the above-$6$ GHz network, however, we jointly consider the locations of both interferers, blockages, and directional signal transmissions. The results allow the prediction of $\SIR$ levels at the secondary RXs, as shown in Fig.~\ref{Fig:SIRcompare}. Furthermore, we show that the interference levels at the two locations become independent when {the measured interference level is high} and/or when many blockages are in place or beamwidth of the system is narrow (again, in the case of the above-$6$ GHz network). Finally, we show that the optimal access probability is proportional to the OP under the specific condition on the secondary users' density.
\item {\bf SaP framework optimization}: 
The SaP framework is designed to maximize the \emph{area spectral efficiency} (ASE) of the secondary network, defined as the sum of the throughput of secondary TXs within an unit area, in terms of the minimum \emph{signal-to-interference ratio} ($\SIR$) threshold required to decode the secondary signal. 
Note that important design issues are on the relation between the access threshold and the minimum decoding target.
For example, a small access threshold increases access probability but reduces transmission success probability due to higher aggregate interference. 
To compensate for such loss, we can decrease the $\SIR$ target but this reduces link quality. 
To deal with this tradeoff, we use a linear mapping function with an access threshold being identical to the decoding threshold whose near-to-optimal ASE is verified by analysis and simulation.  
Some insightful observations are provided by the following optimization. 
First, the optimal mapping function plays a role to make the density of concurrent transmitting TXs constant regardless of the secondary TX density. 
Second, the optimal target $\SIR$ increases with the building density and the primary users' beamwidth. 
\item {\bf Testbed and numerical verification}: We performed testbed experiments using \emph{universal software radio peripheral} (USRP) and MATLAB simulations to verify the accuracy of our results. A real-world building geography of three cities is also reflected in the MATLAB simulations for the practical viability of the outcomes. For more reliability, the simulations include comprehensive functions such as hidden/exposed node problems and outage condition. This helps evaluate the derived spatial correlations and the significant ASE improvements afforded by SaP.
\end{itemize}


\vspace {-15px}
\section{System Model}
\vspace {-5px}

\subsection{Network Model}\label{Sec:Network_Model}
\vspace {-5px}
Consider a CR network where subscript $k\in\{1,2\}$ denotes primary and secondary networks, respectively. It is assumed that the activities of primary networks are not changed during one slot\footnote{It is reasonable in the sense that the ON-OFF interval of conventional macro base stations, ranging from $5$ to $15$ minutes \cite{nokia,ericsson} is relatively larger than the time slot. It is interesting to consider more sophisticated primary activity models such as a hidden Markov model \cite{surveyACTIVITYMODEL}, a multilayer perceptron-based model \cite{SPECTRUM_PRIDICTION}, and a partially observable Markov decision process \cite{CRMACsurvey}, which is out of scope of this work.} including sensing and predicting durations. The secondary TXs access the spectrum with an underlay approach, in which the secondary users are permitted to use the medium only when the primary users' reliable communications are ensured.
The coordinates of the $k$-th network TXs follow a homogeneous Poisson point process (PPP) $\Phi_k$ with density $\lambda_k$. Processes $\Phi_1$ and $\Phi_2$ are mutually independent. Each secondary TX is assumed to have a paired RX at a distance of $d$. 
We separately consider below- and above-$6$ GHz frequency spectra depending on the blockage and directional transmission effects. 
As mentioned, the blockage effect can be ignored in the below-$6$ GHz scenario but not in the above-$6$ GHz scenario. 
Blockage is modeled as follows. 
The distribution of blockage central point process $\Phi_b$ is stationary and isotropic in terms of density~$\lambda_b$. 
Each blockage follows a Boolean model featuring rectangles of average length $\L$ and width $\W$. To describe the blockage effect, the LOS ball of \cite{Bai14} is adopted, where LOS is guaranteed if the distance is no more than the average LOS, denoted by $R_L$, given as: 
\vspace {-5px}
\begin{align}\label{AverageLOSDist}
R_L=\frac{\pi\sqrt{2 \exp\(-\lambda_b \L \W\)}}{2 \lambda_b \(\L+\W\)}.
\end{align}
For simplicity, we assume that the location of the typical RX does not overlap with a blockage.
\vspace {-30px}
\subsection{Channel Model}
\vspace{-5px}
Each TX in the $k$-th network transmits with a power $P_k$. 
The transmitted signal experiences Rayleigh fading\footnote{Although the Rayleigh model may not fit well with the real mmW environment because mmW signals are LOS-dependent, use of this model simplifies the analytical expressions and provides a lower bound for the downlink rate under conditions of Nakagami fading (see \cite{JH15,Elshaer16}). Furthermore, measurements show that small-scale fading has relatively little influence on mmW communications \cite{mmw_intro1}, rendering the use of the Rayleigh model.} with a mean of unity appropriate (i.e., $h \sim \exp(1)$). 
The primary and secondary networks share the same spectrum, causing co-channel interference. By Slyvnak's theorem \cite{kendall}, the $\SIR$ at a typical $k$-th network RX located at the origin is represented as:
\vspace {-5px}
\begin{align}\label{Eq:SIR_Definition}
\SIR_k=\frac{ P_k h^{(0)} \eta_{r^{(0)}}}{\sum_{j=1}^{2}{\sum_{i \in \widetilde{\Phi}_j} { P_j \phi_j^{(i)} h^{(i)}_{j} \eta_{r_j^{(i)}} }}},
\end{align}where $r^{(0)}$ and $r_j^{(i)}$ respectively represent the distances to a typical TX and TX $i$ of the $j$-th network, the path-loss function $\eta_r$ is given as:\vspace{-5px}
\begin{align}
\eta_r =
\begin{cases}
{r}^{-\alpha}, & {\textrm{in below-$6$ GHz,}}\\
\boldsymbol{1}_{r}{r}^{-\alpha}, & {\textrm{in above-$6$ GHz,}}
\end{cases}
\end{align} 
with the indicator function $\boldsymbol{1}_r$ returning unity if $r\leq R_L$, and the active interferers of the $j$-th network form a PPP given by $\widetilde{\Phi}_j$ with density $\lambda_j\frac{\omega_j}{2\pi}$ with the corresponding beamwidth $\omega_j$, resulting from thinning $\Phi_j$.
The path-loss exponent is $\alpha >2$.
The access indicator ${\phi}_j^{(i)}$ is a binary variable of $1$ if the $i$-th nearest TX of the $j$-th network accesses the spectrum, and $0$ otherwise. 
Every primary TX is assumed to transmit constantly (i.e., $\phi_1^{(i)}=1, \forall i$), whereas secondary TX transmit in accordance with the MAC design (as explained below). 
\vspace {-10px}
\section{Sense-and-Predict: Operation, Optimality and Problem Formulation}\label{sec_3}
\vspace {-5px}
In this section, we introduce our proposed MAC, termed SaP where a secondary TX transmits data using OP-based random access, and we discuss on its optimality. Given the framework, the problem of maximizing the area spectral efficiency is formulated.
\vspace {-15px}

\subsection{Sense-and-predict Operation}
The operation of SaP is elaborated as follows. Time is slotted; each slot includes sensing and transmission periods synchronized among secondary TXs. 
During a sensing period, every secondary TX measures the aggregate interference, 
denoted by $I$, from the primary TXs. Assume that the sensing period is sufficient to enable accurate measurement of~$I$. 
Each secondary TX predicts the primary interference at its paired RX separated by the distance of $d$ conditioned on~$I$. 
Specifically, assuming that the other secondary TXs are silent, ($\phi_2^{(i)}$ in \eqref{Eq:SIR_Definition} is zero for all~$i$), it calculates the conditional coverage probability of the received $\SIR_2$ larger than a threshold $\theta$ given a sensed level of interference $I$: \vspace{-5px}
\begin{align}\label{Eq:DefinitionOP}
\mathsf{P}(\theta| I)=\Pr[\SIR_2 \geq \theta | I, \phi_2^{(i)}=0, \forall i],
\end{align}
and this is defined as the OP. 
The OP value $\mathsf{P}(\theta| I)$ determines the random access probability of a secondary TX using an one-to-one mapping function $\mathcal{F}(x)$ to transform the level of OP to an access probability, optimization of which will be discussed below. During a transmission period, each secondary TX accesses the medium with a probability of $\mathcal{F}(\mathsf{P}(\theta| I))$,~namely, \vspace{-5px}
\begin{align}\label{Eq:AccessIndicator}
\phi_2 = \l\{\begin{aligned} 
&1, && \text{w.p}\ \mathcal{F}(\mathsf{P}(\theta| I)), \\
&0, && \text{w.p}\ 1 - \mathcal{F}(\mathsf{P}(\theta| I)). 
\end{aligned}
\r.
\end{align}
When a secondary TX decides to transmit (i.e. $\phi_2=1$), its communication link setup process is performed including a channel estimation and synchronization between the pair of TX and RX.

\vspace {-15px}

\subsection{Optimality of Opportunistic Probability Based Random Access}\label{subsection:optimality}
Given the OP value $\mathsf{P}$ \eqref{Eq:DefinitionOP}, this subsection aims at optimizing the mapping function $\mathcal{F}(x)$ to maximize ASE of secondary network $\lambda_2\mathsf{E}_I \[\mathcal{F}\left( \mathsf{P}(\theta| I ) \right) \cdot \Pr[\SIR_2 \geq \beta|I]\] \cdot \ln(1+\beta)$, where $\mathcal{F}(\mathsf{P}(\theta|I))$ is the access probability specified in \eqref{Eq:AccessIndicator} and $ \Pr[\SIR_2 \geq \beta]$ is the corresponding transmission success probability with the target decoding threshold $\beta$. Using an approach similar to the well-known method to derive network success probabilities using stochastic geometry (see e.g., $\SIR$ coverage in~\cite{Andrews11}), we obtain the following  ASE of the secondary network:
	\begin{align}
	\mathcal{A}&=\lambda_2\mathsf{E}_I \[\mathcal{F}\left( \mathsf{P}(\theta| I ) \right) \cdot \Pr[\SIR_2 \geq \beta|I]\] \cdot \ln(1+\beta)\nonumber\\
	&\stackrel{(a)}=\lambda_2\ln(1+\beta) \int_0^\infty \underbrace{\mathcal{F}\left( \mathsf{P}(\theta| \ell ) \right)}_{\textrm{Access probability}}
	\underbrace{\mathsf{P}(\beta| \ell )}_{\text{$\SIR$ coverage regarding}\atop \text{primary interference}}		\underbrace{\exp \left( { - \pi {\lambda _2}d^2 \rho_0(\beta,\infty) } \right)^{\E_{I}\left[ {\mathcal{F}\left( \mathsf{P}(\theta| I ) \right)} \right]}}_{\text{$\SIR$ coverage regarding secondary interference}} f_{I}\left( \ell \right)d{\ell}\nonumber\\
	& =\lambda_2\ln(1+\beta) {A^{\E_{I}\left[ {\mathcal{F}\left(\mathsf{P}(\theta|I)\right)} \right]}}\int_0^\infty {{\mathcal{F}\left(\mathsf{P}(\theta|\ell)\right)}\mathsf{P}(\beta|\ell) f_{I}\left( \ell \right)d\ell}, \label{opt_prob_sircov}
	\end{align}
where $\rho_0(\beta,t):=\beta^{\frac{2}{\alpha}}\int_0^t {\frac{du}{1+u^{\frac{\alpha}{2}}}}$ and $A=\exp \left( { - \pi {\lambda _2}d^2 \rho_0(\beta,\infty) } \right)$. The terms $f_{I}\left( \ell \right)$ and ${\E_{I}\left[ {\mathcal{F}\left(\mathsf{P}(\theta|I)\right)} \right]}$ represent the \emph{probability density function} (PDF) of measured interference $I$ and the expected value of mapping function $\mathcal{F}\(\mathsf{P}(\theta|I)\)$, respectively. Step (\emph{a}) follows from the fact that the $\SIR$ coverage of multi-tier networks can be decomposed into the $\SIR$ coverage regarding interference of each tier, when the processes of all tiers are independent to each other~\cite{offloading13}.

For simplicity, consider a linear mapping function with an access threshold  $\theta$ being identical to~$\beta$, namely,
 $\mathcal{F}(x)=c \cdot x$ and $\theta=\beta$ where $c$ is a constant. The optimality on this linear mapping function will be discussed in Remark \ref{remark:OptimalMapping}. Recalling that the output of the mapping function is no more than one, i.e., ${\mathcal{F}(x)}\leq1$, plugging this mapping function into \eqref{opt_prob_sircov} and differentiating it in terms of $c$ provide the optimal scaling $c^*$ maximizing ASE $\mathcal{A}$ as  in the following proposition.  
 
	\vspace{-10px}
	\begin{proposition}[Optimal Scaling Factor] \emph{Consider a linear mapping function $\mathcal{F}(x)=c \cdot x$ with~$\theta=\beta$. When the decoding threshold $\beta$ is given, the optimal scaling factor maximizing ASE $\mathcal{A}$ \eqref{opt_prob_sircov}, denoted by $c^*$, is given as 
			\begin{align}
			c^*=\frac{1}{{\pi {\lambda _2}d^2 \rho_0(\beta,\infty) }\mathsf{E}_{I}\left[ \mathsf{P}(\beta|I)\right]}, \label{opt_prob_cond}
			\end{align} 
			if $c^*\leq 1$. 			Here,  $\rho_0(\beta,t)$ is specified in \eqref{opt_prob_sircov}.
			 } \label{prop_opt}
	\end{proposition}
\vspace{-20px}
\begin{remark}[Optimality on the Linear Mapping Function]\label{remark:OptimalMapping}\emph{Among the set of entire mapping functions, the proposed linear mapping function in Proposition 1 is  a sub-optimal solution maximizing the ASE $\mathcal{A}$ (6) due to the following step-by-step justification.
\begin{enumerate}
\item Using H{\"o}lder's inequality, the upper bound of ASE $\mathcal{A}$ \eqref{opt_prob_sircov} is given as\vspace{-5px}
	\begin{align}
	\mathcal{A} \le \mathcal{A}_u= \lambda_2\ln(1+\beta) {A^{\E_{I}\left[ {\mathcal{F}\left(\mathsf{P}(\theta|I)\right)} \right]}\sqrt {\int_0^\infty {{{\mathcal{F}^2}\left(\mathsf{P}(\theta| \ell)\right)}f_{I}\left( \ell \right)d\ell} } }\sqrt {\int_0^\infty {{\mathsf{P}^2(\beta|\ell)}f_{I}\left( \ell \right)d\ell} }, \label{afterholder}
	\end{align}
	where the equality holds if ${\mathcal{F}\left(\mathsf{P}(\theta|\ell)\right)}$ and $\mathsf{P}(\beta|\ell)$ for all $I=\ell$ are linearly dependent, namely, ${\mathcal{F}\left(\mathsf{P}(\theta|\ell)\right)}\propto\mathsf{P}(\beta|\ell)$. 
\item Using Jensen's approximation, we can approximate the upper bound above as 
\begin{align} \label{approximated_upper_bound_ase}
\mathcal{A}_u\approx \lambda_2\ln(1+\beta) {A^{\E_{I}\left[ {\mathcal{F}\left(\mathsf{P}(\theta| I)\right)} \right]} }\E_{I}\left[ {\mathcal{F}\left(\mathsf{P}(\theta| I)\right)} \right]\sqrt {\int_0^\infty {{\mathsf{P}^2(\beta|\ell)}f_{I}\left( \ell \right)d\ell} },
\end{align}
which is a function of the expected mapping function $\E_{I}\left[ {\mathcal{F}\left(\mathsf{P}(\theta| I)\right)} \right]$.
\item Consider a feasible subset of $\{\mathcal{F}, \theta\}$ of which the expected mapping function is identical to a constant $a$. Given the decoding threshold $\beta$, this set shares the same approximated upper bound~\eqref{approximated_upper_bound_ase}. In other words, the solution holding the equality in \eqref{opt_prob_sircov} can be an optimal solution to maximize~\eqref{approximated_upper_bound_ase}, e.g., $\theta=\beta$ and $\mathcal{F}(x)=c \cdot x$ with a constant $c$ determined by $a$. Since the entire set of $\{\mathcal{F}, \theta\}$ is exclusively covered by the whole subsets with different $a$, we can conclude that the structure of  the linear mapping function with $\theta=\beta$ always reaches the approximated upper bound of ASE.
\end{enumerate}
In spite of its sub-optimality, its near-to-optimal performance is well-verified by simulation in Sec.~\ref{Performance Evaluations}. For brevity, we hereafter use the target threshold $\beta$ as the access threshold $\theta$.}
\end{remark}

\vspace{-20px}
\begin{remark}\emph{The constant $c^*$ \eqref{opt_prob_cond} has a twofold effect. The first is to determine the feasible condition where Proposition \ref{prop_opt} holds, which is valid  when the density of the secondary TXs $\lambda_2$ is high. It affords a good match to the massive access required by the cognitive IoT \cite{5GMagazine},  one of the main target applications. The second is to scale down the access probability from the OP $\mathsf{P}(\beta| I)$ \eqref{Eq:DefinitionOP}, equivalent to the coverage probability when the other secondary users are silent. It guarantees the target coverage probability $\Pr[\SIR_2 \geq \beta]$  under the concurrent transmissions of multiple secondary users.}
\end{remark}\vspace {-10px}

\vspace {-15px}
\subsection{Problem Formulation}
\vspace {-5px}
We seek to maximize the ASE $\mathcal{A}$ defined as the sum of the data transmission rates of the secondary RXs per unit bandwidth in an unit area \cite{JH15}. To this end, the following problem is formulated: \vspace{-5px}
\begin{align}\label{Eq:ASE_cond} \tag{P1}
&\underset{\beta}{\text{max}}\ \mathcal{A}= \lambda_2 \mathsf{E}_I \[c^*\mathsf{P}(\beta| I)\cdot\Pr[\SIR_2\geq\beta | I]  \] \ln(1+\beta), \nonumber\\
&\ \ \text{s.t. }\ \Pr[\SIR_1 < \gamma] \leq \tau, \nonumber
\end{align}
where the constraint represents the primary protection requirement\footnote{This probabilistic primary user protection is widely used in practice, for example, to protect a satellite system as a primary network several principles are recommended in \cite{ITU-R}. On the other hand, it may cause the failures of \emph{ultra-reliable and low-latency communication} (URLLC)~\cite{urllc}. To avoid it, an additional signaling protocol is designed to prevent the secondary TX's access when URLLC packets exist, which is outside the scope of current work.}, such that the $\SIR$ outage probability of a primary user, $\Pr[\SIR_1<\gamma]$ (where $\gamma$ is a target decoding threshold for a successful primary transmission) does not exceed a given constant $\tau$.

\vspace {-10px}
\begin{remark}[Extension to Full Duplex Networks]\label{remark:FullDuplexEffect} \emph{It is interesting to extend the current SaP design based on half-duplex mode into full-duplex mode when both of the secondary TX and RX can measure OP. It allows more nodes to access the medium simultaneously, leading to network-level performance improvement e.g., ASE. On the other hand, it causes increasing aggregate interference, resulting in the degradation of link-level performance e.g., coverage probability. Consideration of this tradeoff is a key for the extension to full-duplex, deserving further investigation in the future. }\label{FullDuplexEffect}
\end{remark}\vspace {-5px}

\vspace{-5px}
\vspace{-5px}
\vspace{-5px}

\section{Sense-and-Predict in below-$6$ GHz spectrum}\label{sec:woblockage}
\vspace {-5px}
In this section, we attempt to derive a tractable form of OP by applying SG. 
Based on the OP, we optimize the SaP framework by solving Problem \ref{Eq:ASE_cond}. 

\vspace {-15px}
\subsection{Opportunistic Probability Analysis} 
\vspace{-5px}
As mentioned, a secondary TX accesses the spectrum with a probability of OP $\mathsf{P}$ that features the spatial interference correlation between TX and RX. The direct derivation of $\mathsf{P}$ is intractable because the secondary TX and RX share common interferers, causing angular correlations. Such correlations violate the requirement for isotropy in point processes, blocking the use of PPP techniques such as Campbell's theorem~\cite{Haenggi13}. To address this issue, we introduce the following assumption:

\begin{figure}
	\centering 
	{\includegraphics[width=9cm]{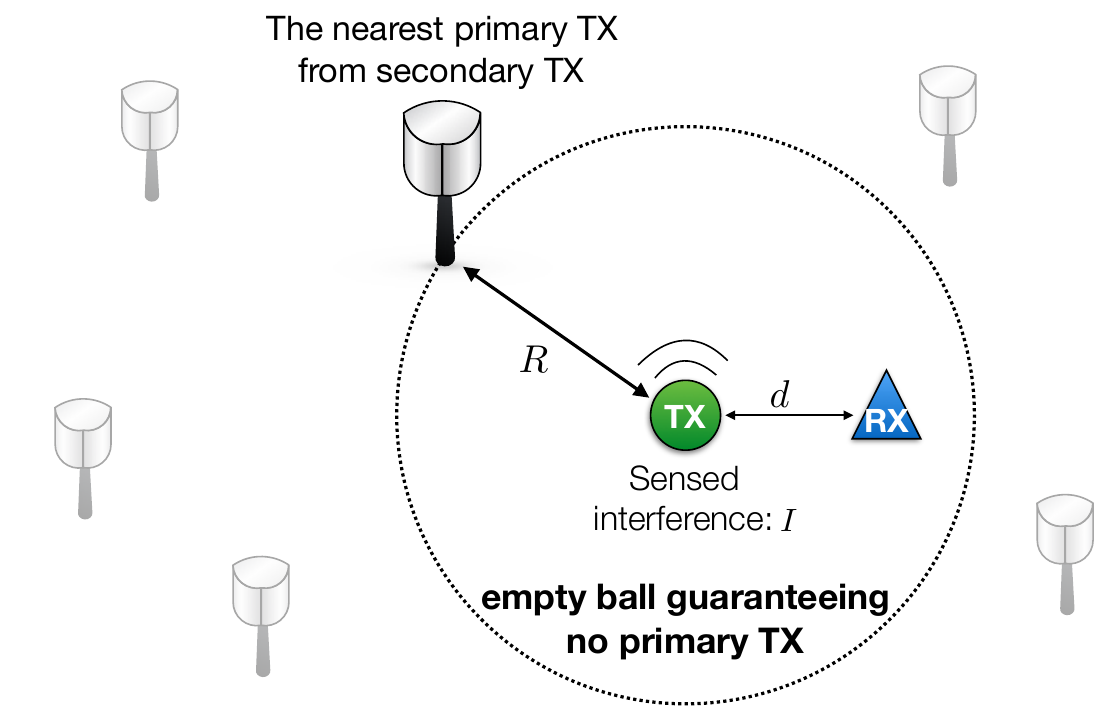}} \vskip-10pt
	\caption{An empty ball with radius $R_I$ determined by the sensed interference level $I$.\label{derivation}}\vspace{-.35in}
\end{figure}
\vspace{-.1in}
\begin{assumption}[Empty Ball]\label{EmptyBallAssumption}
	\emph{
		The nearest primary TX is assumed to be located on a circle with radius $R$ defined as an \emph{empty ball}. In other words, no primary TX exists inside the empty ball, as shown in Fig.~\ref{derivation}. Furthermore, the other primary TXs (located outside the empty ball) are assumed to follow PPP from the perspective of a typical secondary RX.
	}
\end{assumption}
\vspace{-5px}
This assumption \ref{EmptyBallAssumption} allows us to tractably manage the angular correlation problem by decomposing $\mathsf{P}$ into a product of two terms. The first term corresponds to the interference from the nearest primary TX on the empty ball, the angular correlation of which can be expressed using a single trigonometrical function. The second term corresponds to the aggregate interference from the other primary TXs outside the empty ball, the angular correlations of which can be disregarded because of the isotropic property of PPP. The principal result is shown in the following proposition:

\vspace{-10px}
	\begin{proposition}[OP Analysis in Below-$6$ GHz Spectrum]\label{prop_OP1}
		\emph{
			Assuming that the empty ball radius $R$ is known, the OP $\mathsf{P}$ is given as $\mathcal{{P}}(R, d, \beta)$,\vspace{-5px}
			\begin{align}
			&\mathcal{{P}}(R, d, \beta) = \frac{1}{2\pi} \int_{0}^{2\pi}\frac{P_2 d\nu}{P_2+{P_1} d^\alpha \left(R^2-2 d R \cos(\nu) + d^2\right)^{-\frac{\alpha}{2}}} 
			\exp\[-\int_{[R-d]^+}^{\infty}c_0(y,R) dy\]^{\beta^{\frac{2}{\alpha}}}, \label{Eq:CondSIR1} 
			\end{align}\vspace{-5px}
			where $[x]^+=\max[0,{x}]$, $c_0(y,R)=\frac{2\lambda_1{P_1 } d^\alpha y^{-\alpha+1}}{P_2+{P_1 } d^\alpha y^{-\alpha}}{\arccos\left(\frac{{R}^2-d^2-y^2}{2 d y}\right)}$, if ${|R-d|} < y \leq {R+d} $ and $c_0(y,R)= \frac{2\pi\lambda_1{P_1 } d^\alpha y^{-\alpha+1}}{P_2+{P_1 } d^\alpha y^{-\alpha}}$ otherwise.
		}
	\end{proposition}
	\vskip 0pt \noindent {\bf Proof.} 
	See Appendix A. 
	\hfill $\blacksquare$

Proposition \ref{prop_OP1} shows that the closed form of OP $\mathsf{P}$ can be obtained on the condition that the empty ball radius $R$ is known.
Unfortunately, determination of the exact $R$ is impossible because many types of interferer distributions can yield the same aggregate interference. Alternatively, we approximate $\mathsf{P}$ by plugging the conditional expectation of $R$ given the $I$, denoted by $R_I$, into $R$ in \eqref{Eq:CondSIR1}, namely:\vspace{-5px}
\begin{align}
\mathsf{P} \approx\mathcal{{P}}(R_I, d, \beta), \label{OP_ri} \vspace{-10px}
\end{align}
where $R_I$ is obtained from the following problem, with the proof given in Appendix B.\vspace{-5px}
\begin{align}\label{Eq:EmptyBallRadius}
\frac{I}{P_1} {{{R}_I}}^\alpha-\frac{2\pi\lambda_1}{\alpha-2}{{{R}_I}}^2-1=0.
\end{align}
The approximation in \eqref{OP_ri} is verified by simulation to be tight (see Fig.~\ref{fig:CondSIR} (b)). Furthermore, the closed form of $R_I$ is enabled when the path-loss exponent $\alpha=4$,\vspace{-5px}
\begin{align}\label{Eq:ClosedformRI} 
{R}_I = (2I)^{-\frac{1}{2}}\({{\pi\lambda_1 P_1 + \[{\(\pi\lambda_1 P_1\)^2+4P_1 I}\]^{\frac{1}{2}}}}\)^{\frac{1}{2}},
\end{align}
which is explicitly shown to increase with  $\lambda_1$ and $P_1$, and to decrease with $I$. We can simplify $\mathcal{{P}}(R_I, d, \beta)$ in \eqref{OP_ri} using the following asymptotical analysis, enabling us to understand of the relations between network parameters and the resultant OP. 
\vspace {-10px}

\begin{corollary}[Asymptotic OP]\emph{When the expected empty ball radius $R_I$ increases, $\mathcal{{P}}(R_I, d, \beta)$ of \eqref{OP_ri} tends to follow $\mathcal{{P}}(R_I, 0, \beta)$ and finally converges to $1$. 
		On the other hand, when $R_I$ decreases, $\mathcal{{P}}(R_I, d, \beta)$ decreases and converges to the following value, \vspace{-5px}
		\begin{align}\label{Eq:AsymptoticOP}
		\lim_{R_I\rightarrow 0}\mathcal{{P}}(R_I, d, \beta)=\mathcal{{P}}(0, d, \beta)=\frac{P_2 \exp\({-\pi\lambda_1\(\frac{P_1\beta d^\alpha}{P_2}\)^{\frac{2}{\alpha}}\int_0^\infty {\frac{du}{1+u^{\frac{\alpha}{2}}}}}\)}{P_2+P_1\beta},
		\end{align} 
		which is strictly larger than zero and independent of $I$ and $R_I$. 
	}\label{asmyptotic_OP}
\end{corollary}

\vspace {-15px}
\begin{remark}[Effects of Parameters]\label{remark:ParameterEffect} \emph{Some interesting observations arise from \eqref{Eq:ClosedformRI} and Corollary \ref{asmyptotic_OP}. First, even though the same interference level is sensed, the access probability can differ depending on the network parameters, including the primary TX density $\lambda_1$ and the transmission power $P_1$. Second, the difference in interference level between a typical secondary TX and an RX location falls as $R_I$ increases, implying that a smaller $I$ in environments with larger $\lambda_1$ and $P_1$ values allows secondary TXs to access the spectrum more reliably. Lastly, the fact that a non-zero lower bound of OP exists encourages the TX to access the spectrum with a certain probability, even though very large numbers of $I$ are sensed. 
	}\label{ri_parameter}
\end{remark}\vspace {-5px}

\vspace {-15px}
\subsection{Area Spectral Efficiency Maximization}
\label{ase_wo_bl}
	This section deals with the ASE maximization in Problem \ref{Eq:ASE_cond}, requiring an analysis of the term $\E_I\l[c^* \mathsf{P}(\beta| I) \cdot\Pr(\SIR_2\geq\beta | I)\r]$ as a preliminary step.
	Note that the term $\mathsf{P}(\beta| I)$ is given as $\mathcal{P}(R_I, d, \beta)$ in \eqref{Eq:CondSIR1} at an empty ball radius of $R$ but the conditional coverage probability $\Pr(\SIR_2\geq\beta| I=\ell)$ has not yet been derived. 
	We express $\Pr(\SIR_2\geq\beta| I=\ell)$ in the form of~$\mathcal{P}(R, d, \beta)$ in \eqref{Eq:CondSIR1}, and optimize $\beta$ tractably.

	For analytical tractability, we assume that the concurrent secondary TXs are independently thinned by the average access probability, and that the radius of the empty ball $R$ is perfectly estimated, equivalent to the nearest primary TX, the PDF of which is $f_R(r)=2\pi\lambda_1 r e^{-\pi\lambda_1 r^2}$. We can express the average access probability $\overline{\mathcal{P}}(d,\beta)$ in terms of 
	$\mathcal{{P}}(R, d, \beta)$ of \eqref{Eq:CondSIR1} as:\vspace{-5px}
	\begin{align}\label{Eq:ExpectedOP}
	\overline{\mathcal{P}}(d,\beta)=c^* \int_0^\infty {\mathcal{{P}} (r,d,\beta)} f_R(r) dr = \frac{1}{{\pi {\lambda _2}d^2 \rho_0(\beta,\infty) }}.
	\end{align}
	Using \eqref{Eq:ExpectedOP}, we can derive the conditional coverage probability $\Pr(\SIR_2\geq\beta| I=\ell)$ given in Lemma~\ref{lemma_avgOP}. In addition, this enables us to calculate the lower limit of decoding $\SIR$ target preventing outage of a primary user; the lower limit exceeds the threshold $\tau$, as proven in Lemma~\ref{lemma_mintheta}:\vspace{-10px}
	\noindent\begin{lemma}[Secondary Coverage Probability]\emph{ \normalsize
			Given the access threshold $\beta$ and the radius of the empty ball $R$, the probability $\Pr(\SIR_2\geq\beta| I=\ell)$ is given as \vspace{-5px}
			\begin{align}
			\Pr(\SIR_2\geq\beta| I=\ell)=\mathcal{{P}}(R_I, d, \beta) \exp \l(-1\r).
			\end{align}			 \vskip 0pt \noindent
			{\bf Proof.} See Appendix C.
			\hfill $\blacksquare$}\label{lemma_avgOP}
	\end{lemma}
\vspace {-10px}
\vspace{-10px}
	\begin{remark} \emph{ 
			This result shows that the second term of secondary coverage probability is a constant value, erasing the effect of secondary interferers in the secondary coverage probability. This implies that the access probability $c^*\mathcal{{P}}(R_I, d, \beta)$ is consequently linearly dependent on the secondary coverage probability.}
	\end{remark}

\vspace{-10px}
\vspace{-10px}
\noindent\begin{lemma}[Minimum Decoding Target] \emph{To satisfy the constraint of \ref{Eq:ASE_cond}, the decoding target $\beta$ should be no less than ${\beta_{\min}}$, which is given as: \vspace{-5px}
		{\begin{align}
			\beta_{\min}=\(\frac{{P_1}^{\frac{2}{\alpha}} \rho_0(\gamma,\infty) (1-\tau) }{\pi d^2 \int_0^\infty\frac{du}{1+u^\frac{\alpha}{2}} \lambda_1 {P_2}^{\frac{2}{\alpha}} \[\tau\!+\! \rho(\gamma,\infty)\tau-\rho(\gamma,\infty)\]}\)^{\frac{\alpha}{2}},
			\end{align}}\
		where $\rho(a,t):=a^{\frac{2}{\alpha}}\int_{a^{-\frac{2}{\alpha}}}^{t}\frac{du}{1+u^\frac{\alpha}{2}}$
		\vskip 0pt \noindent
		{\bf Proof.} See Appendix D. 
		\hfill $\blacksquare$}\label{lemma_mintheta}
\end{lemma}
\vspace {-10px}
From Lemmas \ref{lemma_avgOP} and \ref{lemma_mintheta}, \ref{Eq:ASE_cond} is rewritten as\vspace{-5px}
\begin{align}
\underset{\beta\geq{\beta_{\min}}}{ \max}\mathcal{A}= 
\ln(1+\beta) 
\exp\l(-1\r) c^* \lambda_2
\int_0^\infty & \mathcal{{P}}(r,d,\beta) \mathcal{{P}}(r,d,\beta) f_{R}(r) dr.\label{ProblemReformulation}\tag{P2}
\end{align}
\noindent 
Using the mean value theorem, the integral term in \ref{ProblemReformulation} is decomposed as:\vspace{-5px}
	\begin{align}
	\int_0^\infty  \mathcal{{P}}(r,d,\beta) \mathcal{{P}}(r,d,\beta) f_{R}(r) dr
	&=\mathcal{{P}}(s,d,\beta) 	{\int_0^\infty \mathcal{{P}}(r,d,\beta)  f_{R}(r) dr},
	\end{align}
	where $s$ is a positive constant which satisfies $\mathcal{{P}}(s,d,\beta) =\frac{\int_0^\infty \mathcal{{P}}^2(r,d,\beta)  f_{R}(r) dr}{ \int_0^\infty \mathcal{{P}}(r,d,\beta)  f_{R}(r) dr}$.
	By eliminating terms not related to $\beta$, \ref{ProblemReformulation} becomes\vspace{-5px}
	\begin{align}
	\underset{\beta\geq{\beta_{\min}}}{ \max} 
	\ln(1+\beta)  \beta^{-\frac{2}{\alpha}} \exp\[-\int_{[s-d]^+}^{\infty} {c_0(y,s)} dy\] ^{\beta^{\frac{2}{\alpha}}} ,\label{ASE_1_re}
	\end{align}
	where $c_0(y,s)$ is specified in \eqref{Eq:CondSIR1}.
	By differentiating the equation \eqref{ASE_1_re}, we derive the optimal decoding target as in the following proposition.
	\vspace{-10px}
	\begin{proposition}[Optimal Decoding Target in below-$6$ GHz]\emph{When the path-loss exponent $\alpha > 2$, the optimal decoding target $\beta^*$ is represented as:\vspace{-5px}
	\begin{align}\label{optimal_beta}
	\beta^* = \max(\beta_0, \beta_{\min}),
	\end{align}
where $\beta_0$ is a value that satisfies $-2\int_{[s-d]^+}^{\infty} {c_0(y,s)} dy{\beta_0}^{\frac{2}{\alpha}} \ln(1+\beta_0) + \frac{\alpha\beta_0}{1+\beta_0} - 2\ln(1+\beta_0)=0$ and $\beta_{\min}$ is specified in Lemma \ref{lemma_mintheta}.
\vskip 5pt \noindent
{\bf Proof.} 
See Appendix E.
\hfill $\blacksquare$}\label{prop_optimal_beta}
\end{proposition}
\vspace {-10px}


Combining Lemmas \ref{lemma_avgOP}, \ref{lemma_mintheta} and Proposition \ref{prop_optimal_beta}, one can infer that there is an optimal density for concurrent secondary transmissions. The following remark specifies this observation.
\vspace {-10px}
\begin{remark}[Optimal Concurrent Transmitting TX Density] \emph{This result shows that an optimal density for concurrent secondary transmissions in fact exists: ${\Lambda}_2^*=\lambda_2 {\overline{\mathcal{P}}(d, \beta^*)}= \frac{1}{\pi d^2 \rho_0(\beta^*,\infty) }$, yielding the expression of the relationships between parameters. First, as $\lambda_2$ increases, the optimal access threshold $\beta^*$ should increase to retain the optimal density $\Lambda_2^*$. Second, when the overall OP values are high (i.e. when $\mathcal{{P}}(c,d,\beta) \sim 1$), the optimal decoding target $\beta^*$ satisfies:~$\frac{\beta^*}{(1+\beta^*)\ln(1+\beta^*)}=\frac{2}{\alpha}$, implying that $\beta^*$ decreases with the path-loss exponent $\alpha$.}\end{remark}



\vspace{-15px}

\section{Sense-and-Predict in above-$6$ GHz spectrum}

In this section, we specify the OP analysis to a case with blockages, then extend the OP analysis by considering the directional transmissions to make it more suitable for the mmW scenario. We henceforth maximize the corresponding ASE, providing the target $\SIR$ for decoding.

\vspace {-10px}
\subsection{Opportunistic Probability Analysis with Blockage Effects}

\begin{figure*}   
\centering
  \subfigure[Actual LOS regions of a secondary pair]{\centering
   \includegraphics[width=7cm]{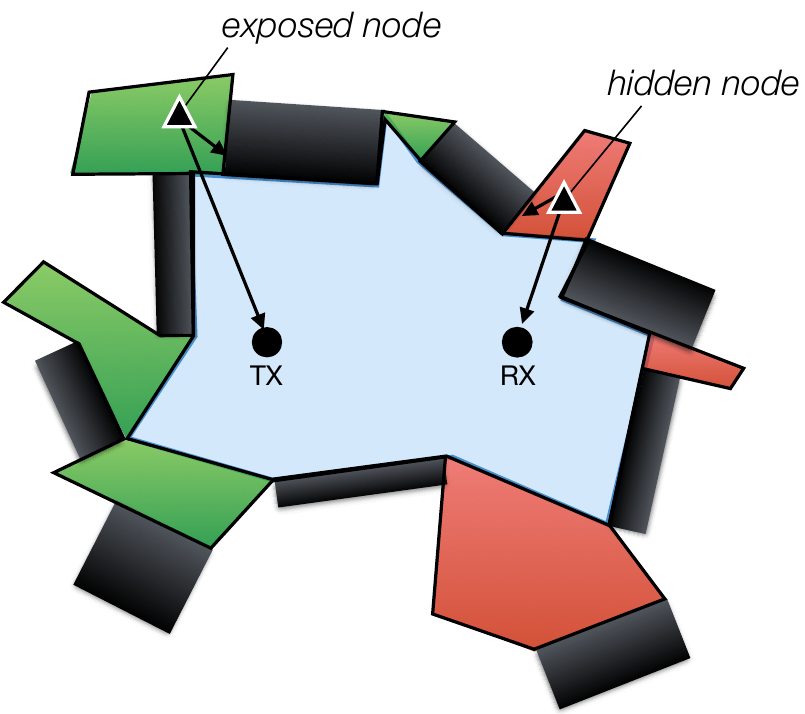} } 
  \subfigure[Approximated joint unblocked region of a secondary pair]{
   \includegraphics[width=7cm]{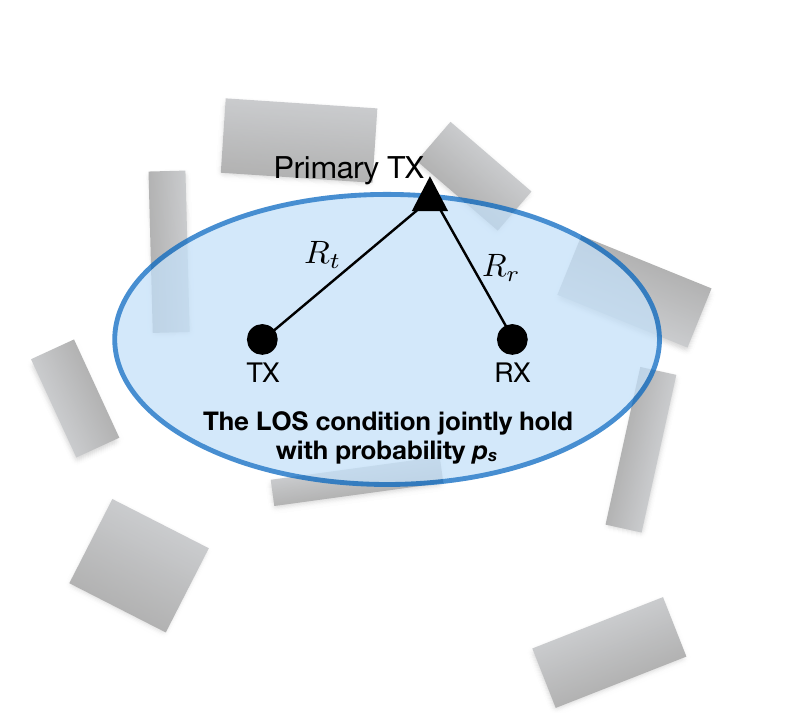}} \vskip-10pt
\caption{Joint unblocked region of a secondary pair: assuming the LOS probabilities of two links are independent, the joint unblocked probability of a primary TX located on the ellipse with the secondary pair at two focal points is identical. \label{Fig:jointlos}}\vspace{-.4in}
\end{figure*} 

The analysis of OP incorporating the blockage effect is of prime concern in this subsection. Because of blockages, the link from a primary TX to either a secondary TX or a paired RX can be blocked, causing exposed or hidden node problems (see Fig.~\ref{Fig:jointlos} (a)). To incorporate these problems into OP, we define an \emph{unblocked probability}, denoted by $p$; the probability that the link between a primary TX and a typical secondary TX or RX is unblocked. In \cite{Bai14}, this is expressed in terms of the distance $R$ as \vspace{-5px}
\begin{align}
p(R)=\exp\left[-\frac{2\lambda_b \(\L+\W\)}{\pi}R\right],
\end{align}
where the blockage-related parameters $\lambda_b$, $\L$ and $\W$ are described in Sec.~\ref{Sec:Network_Model}.
For simplicity, we ignore the case in which the two links are blocked by the same blockage. A joint unblocked probability of two links, denoted by $p_{\mathrm{u}}$, is then represented by the product of the individual unblocked probabilities of the secondary TX and RX as:\vspace{-5px} 
\begin{align}\label{Eq:JointLOSProb}
p_{\mathrm{u}}(R_t, R_r) =p(R_t)\cdot p(R_r)=\exp\[-\frac{2\lambda_b \(\L+\W\)}{\pi}\(R_t+R_r\)\],
\end{align}
where $R_t$ and $R_r$ denote the distance from a primary TX to the typical secondary TX and its RX, respectively. It is observed that
$p_{\mathrm{u}}$ in \eqref{Eq:JointLOSProb} is not changed if the sum of the distances between the two links, $R_t+R_r$, is identical. Based on the intuition, we introduce the following assumption:

\vspace {-10px}
\begin{assumption}[Joint Unblocked Region]\label{jointlos_prob}
\emph{Both links from a primary TX to a typical secondary TX and the RX are assumed to be unblocked if the sum of their distance $R_t+R_r$ is no more than $L$, which is determined by blockage-related parameters\footnote{ The distance $L$ can be obtained via data curve-fitting based on real geometric data, as will be described in Section~VI-C.} (e.g., $\lambda_b$, $\L$ and $\W$). The geometry of the joint unblocked area is an ellipse, the two focal points of which are the secondary TX and RX locations and the major axis length  $L$. }
\end{assumption}
\vspace {-10px}
{Assumption \ref{jointlos_prob} allows us to quantify the extent of the exposed node problem. For example, consider one primary TX, of which the link to a typical secondary TX is unblocked. If the primary TX is within the joint unblocked region, it will also be unblocked to the paired secondary RX. Otherwise, it does not interfere with the RX, thereby causing an exposed node problem. Note that it is straightforward to feature a hidden node problem by considering additional interference, which is unlikely to be sensed by the TX, but will be sensed by the RX. The principal result is shown in the following proposition.}

\vspace{-.1in}
 \begin{proposition}[OP Analysis with Blockage Effect]\emph{
Assume that the empty ball radius $R$ and the axis length $L$ are given. 
Under the LOS condition between a typical secondary TX and RX, i.e., $d\leq R_L$ where $R_L$ is specified in \eqref{AverageLOSDist}, the OP $\mathsf{P}$ is given by $\mathcal{P}(R,L,d,\beta)$,\vspace{-5px}
{
\begin{align}\nonumber
\mathcal{P}(R,L,d, \beta) &=\exp\[-\int_{[R-d]^+}^{R_L}c_2(y,R) dy\]^{\beta^{\frac{2}{\alpha}}}\cdot \\
& 
\begin{cases}  
\int_{0}^{\pi}\frac{P_2 \pi^{-1} d\nu }{P_2+{P_1 } d^\alpha \left(R^2-2 d R \cos(\nu) + d^2\right)^{-\frac{\alpha}{2}}}& {\textrm{if ${R<\frac{L}{2}-\frac{d}{2}}$,}} \\
\int_{0}^{u}\frac{P_2 \pi^{-1} d\nu }{P_2+{P_1 } d^\alpha \left(R^2-2 d R \cos(\nu) + d^2\right)^{-\frac{\alpha}{2}}} +\frac{\pi-u}{\pi} & {\textrm{if ${\frac{L}{2}-\frac{d}{2}\leq R<\frac{L}{2}+\frac{d}{2}} $,}} \\
1& {\textrm{if ${\frac{L}{2}+\frac{d}{2} \leq R}$,}} 
\end{cases}
 \label{Eq:CondSIR2} 
 \end{align}}\normalsize
 where $c_2(y,R)=  \frac{2\lambda_1{P_1 } d^\alpha y^{-\alpha+1}}{P_2+{P_1 } d^\alpha y^{-\alpha}}{\arccos\left(\frac{{R}^2-d^2-y^2}{2 d y}\right)}$ if ${\min\(R_L,|R-d|\)} < y \leq {\min\(R_L,R+d\)} $ and $c_2(y,R)= \frac{2\pi\lambda_1{P_1 } d^\alpha y^{-\alpha+1}}{P_2+{P_1 } d^\alpha y^{-\alpha}}$ otherwise, and $u = \arccos\(\frac{d^2 + 2L R - {L}^2}{2 d R}\)$. It is obvious that $\mathcal{P}(R,L,d, \beta)$ is zero if $d> R_L$.
\vskip 0pt \noindent
{\bf Proof.} 
See Appendix F.
 \hfill $\blacksquare$}\label{prop_OP2}
\end{proposition}
\vspace {-10px}
\begin{figure*}   
\centering
  \subfigure[Hidden node problem]{
   \includegraphics[width=6.8cm]{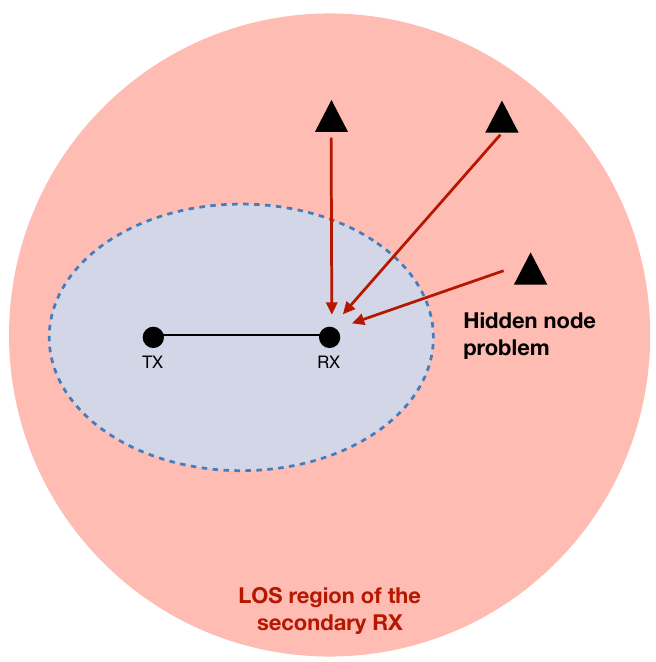}}
   \qquad
    \subfigure[Exposed node problem]{
   \includegraphics[width=6.8cm]{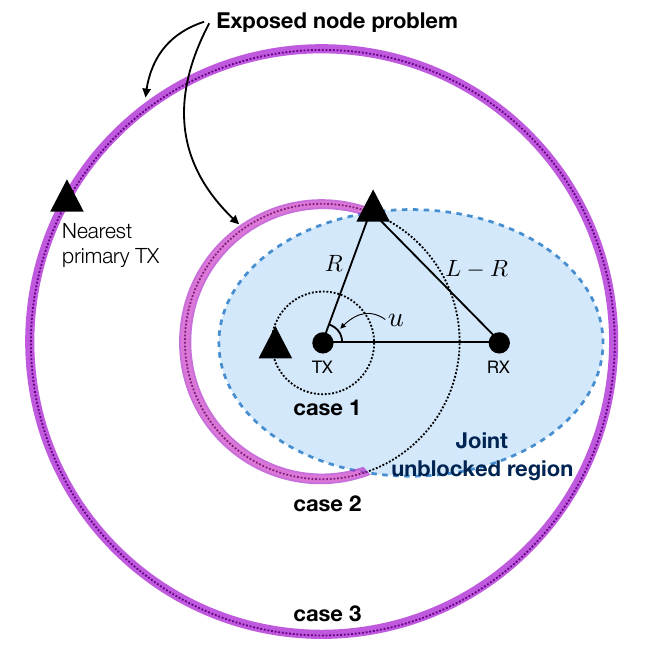} \label{fig:blockage_exp}} \vskip-10pt
\caption{Joint unblocked region and exposed/hidden node problem.\label{fig:blockage_t1}}\vspace{-.4in}
\end{figure*} 
The first part of \eqref{Eq:CondSIR2} represents the $\SIR$ coverage considering the aggregate interference from all primary TXs except the nearest TX, 
including TXs with hidden node problems (see Fig.~\ref{fig:blockage_t1}(a)). The second part represents the $\SIR$ coverage considering interference from the nearest primary TX imposed on the secondary TX, by capturing whether or not this is an exposed node with respect to the secondary RX (see Fig.~\ref{fig:blockage_t1}(b)). Specifically, we consider the following three cases depending on the distance between the secondary TX and the nearest primary TX $R$, namely: 
{\bf Case 1)} no exposed node problem occurs when $R<\frac{L}{2}-\frac{d}{2}$; {\bf Case 2)} a partial problem occurs when $\frac{L}{2}-\frac{d}{2}\leq R<\frac{L}{2}+\frac{d}{2}$; and, {\bf Case 3)}\footnote{In Case $3$, the nearest primary TX could be regarded as an exposed node even though it is closer to the secondary RX than is the secondary TX, creating an error in the OP analysis. Fortunately, such an error is marginal because the interference level of Case 3 is quite small, and the resultant OP $\mathcal{P}(R,L,d, \beta)$ is nearly unity despite the error.} a problem always occurs when $\frac{L}{2}+\frac{d}{2}\leq R$. The possibility that the nearest primary TX causes an exposed node problem decreases as the distance $R$ increases, equivalent to low-level interference $I$.
\vspace {-10px}
\begin{remark}[Effect of Blockages on OP] \emph{As blockages become larger and/or denser, the axis length $L$ decreases and the second part of the OP \eqref{Eq:CondSIR2} becomes close to unity. This implies that when blockage size or density is high, the correlation between interference at the secondary TX and RX decreases (i.e., the measured interference $I$ has less impact on the OP). However, as the blockage size or density decreases, the axis length $L \to \infty$, and the LOS distance $R_L \to \infty$; and the OP with blockage effect \eqref{Eq:CondSIR2} becomes identical to that without blockage effect as \eqref{Eq:CondSIR1}. 
}
\end{remark} \vspace {-10px}

As in the case without blockage, we approximate $\mathsf{P}$ by replacing the exact $R$ with its expected value, namely:\vspace{-15px}
\begin{align}
\mathsf{P} \approx\mathcal{P}({R}_I, L, d, \beta), \label{OP_ri_bl}
\end{align}
where $R_I$ is straightforwardly derived when modifying the upper limit of integration in \eqref{R_I_prop1} from $\infty$ to $R_L$, as follows.\vspace{-5px}
\begin{align}\label{Eq:EmptyBallRadius_1}
\(\frac{I}{P_1} + \frac{2\pi\lambda_1}{\alpha-2} {R_L}^{2-\alpha} \) {{R_I}}^\alpha-\frac{2\pi\lambda_1}{\alpha-2}{{R_I}}^2-1=0.
\end{align}

Furthermore, the distance $R_I$ is derived as a closed-formula as in the following cases according to the path-loss exponent $\alpha$,\vspace{-10px}
\begin{align}\label{R_I_alpha2}
\lim_{\alpha\to {2}+} R_I &= \[\frac{I}{P_1}-2\pi\lambda_1\log\(R_L\)\]^{-\frac{1}{2}}, \\
\lim_{\alpha\to 4} R_I &= {{{\({{\pi\lambda_1 P_1 + \[{\(\pi\lambda_1 P_1\)^2+4P_1 \(I+\pi\lambda_1P_1 {R_L}^{-2}\)}\]}}\)^{\frac{1}{4}}}{\[2\(I+\pi\lambda_1P_1 {R_L}^{-2}\)\]}^{-\frac{1}{2}}}}.
\end{align}
Note that Remark \ref{remark:ParameterEffect} is still valid in the case with blockage effect.



\begin{figure*}
	\centering
	\hfill
	\begin{minipage}[b]{0.45\textwidth}
  		\centering
  		\includegraphics[width=.95\linewidth]{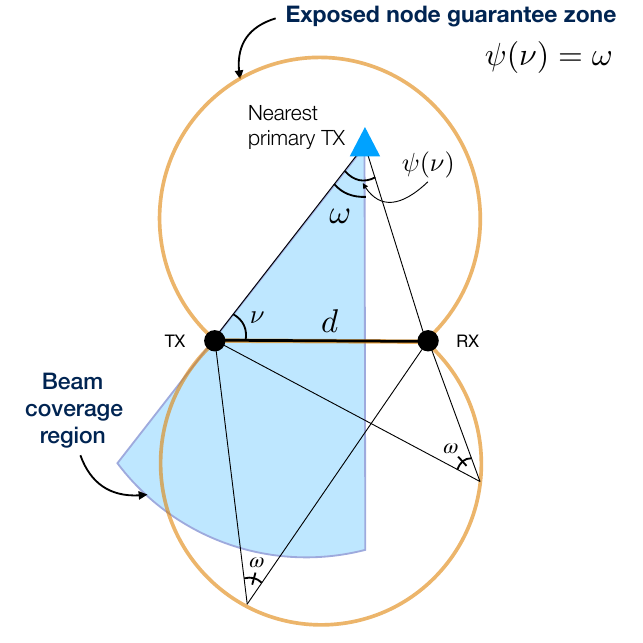}
  		\caption{Occurrence of exposed node and exposed node guarantee zone caused by directional signal transmissions.}
  		\label{fig:directional_exposed_1}
	\end{minipage}%
	\hfill
	\begin{minipage}[b]{0.45\textwidth}
  		\centering
  		\includegraphics[width=.95\linewidth]{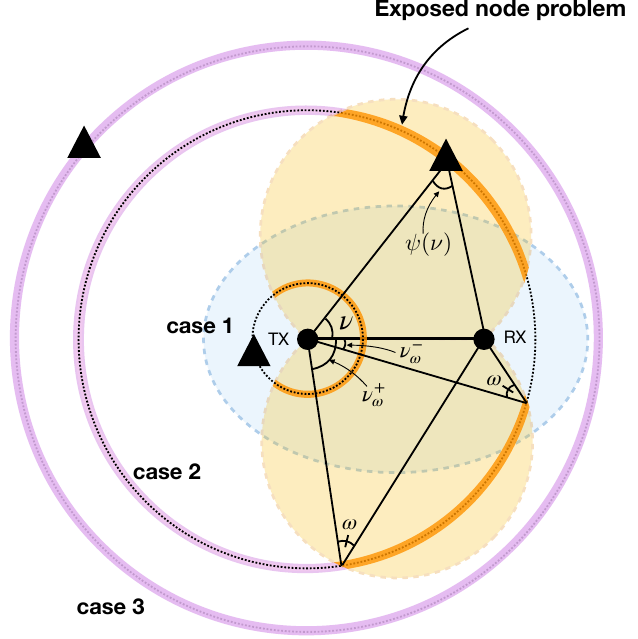}
  		\caption{Different cases of exposed node problem for directional transmissions.}
		\label{fig:directional_exposed_2}
	\end{minipage}
	\hfill
	\vspace{-.4in}
\end{figure*}
\vspace{-.2in}
\subsection{Opportunistic Probability Analysis with Blockage and Directional Transmissions Effects}
This subsection aims at extending the preceding OP analysis by adding a new feature, the directional signal transmission, which is vital in mmW scenario. Besides blockages, the beam direction results in more frequent exposed node problems because a primary TX may not be a common interferer to the secondary TX and RX. As illustrated in Fig.~\ref{fig:directional_exposed_1}, the exposed node problem occurs when the primary TX does interfere with the secondary TX, while it does not interfere with the secondary RX. Consider that each primary TX randomly forms a directional beam with identical beamwidth denoted by $\omega$, which is smaller than $\pi$. One can infer a circle with a circumference angle $\omega$ and a chord whose distance is $d$ (see lower circle in Fig.~\ref{fig:directional_exposed_1}). The primary TX in the circle always causes the exposed node problem. Meanwhile, the primary TX outside of the circle interferes with secondary users simultaneously depending on its angular location. It is shown at the upper triangle in Fig.~\ref{fig:directional_exposed_1} that the interior angle at the primary TX denoted by $\psi$ depends on that at the secondary TX denoted by $\nu$. More specifically, we define a \emph{common interfering probability}, denoted by $p_c$, as the probability that the secondary TX and RX are inside the beam coverage region from the nearest primary interferer as follows:\vspace{-5px}
\begin{align} p_c(\nu)= \begin{cases} 0, \qquad\qquad\qquad&\text{if }\psi(\nu)>\omega, \\
\dfrac{\omega-\psi(\nu)}{\omega}, &\text{otherwise},\end{cases} \label{Eq:common_interfering_prob} \vspace{-5px}\end{align} 
where $\psi(\nu)=\arccos\left(\frac{R-d\cos (\nu)}{\sqrt{R^2+d^2-2Rd\cos(\nu)}}\right)$. In other words, if the primary TX is in the circle, it is always the exposed node, while it becomes exposed with probability $1-p_c(\nu)=\frac{\psi(\nu)}{\omega}$ otherwise.
Next, the aggregate interference from the other primary TXs is straightforward as all the primary TXs are independently thinned by  $\frac{\omega}{2\pi}$ \cite{JH15}. Combining the two leads to the OP for Above-6 GHz Spectrum specified in the following proposition.
\begin{proposition}[OP Analysis in Above-$6$ GHz Spectrum]\emph{Assume that the empty ball radius $R$ and the axis length $L$ are known and the beamwidth of the network is fixed to $\omega$. Under the LOS condition between a typical secondary TX and RX, i.e., $d\leq R_L$ where $R_L$ is specified in \eqref{AverageLOSDist}, the corresponding OP $\mathsf{P}$ is given as $\mathcal{{P}}(R,L, d, \beta, \omega)$,
			{
				\begin{align}\nonumber
				&\quad\qquad \mathcal{{P}}(R, L, d, \beta, \omega) = \exp\[-\int_{[R-d]^+}^{R_L}c_3(y,R) dy\]^{\beta^{\frac{2}{\alpha}}}\cdot \\
				&\begin{cases}  
				\frac{1}{\pi} \int_{0}^{\pi} \left((1-p_c(\nu))+\frac{ p_c(\nu) P_2 }{P_2 + P_1 d^\alpha \({R}^2 - 2 d {R} \cos(\nu) + d^2\)^{-\frac{\alpha}{2}}}\right)d\nu,& {\textrm{if ${R<\frac{L}{2}-\frac{d}{2}}$,}} \\
				\frac{\pi-\mu}{\pi}+\frac{1}{\pi} \int_{0}^{u} \left((1-p_c(\nu))+\frac{ p_c(\nu) P_2 }{P_2 + P_1 d^\alpha \({R}^2 - 2 d {R} \cos(\nu) + d^2\)^{-\frac{\alpha}{2}}}\right)d\nu,  & {\textrm{if ${\frac{L}{2}-\frac{d}{2}\leq R<\frac{L}{2}+\frac{d}{2}} $,}} \\
				1& {\textrm{if ${\frac{L}{2}+\frac{d}{2} \leq R}$,}} 
				\end{cases}
				\label{Eq:CondSIR3} 
				\end{align}}\normalsize
				 where $c_3(y,R)=  \frac{\lambda_1\omega{P_1 } d^\alpha y^{-\alpha+1}/\pi}{P_2+{P_1 } d^\alpha y^{-\alpha}}{\arccos\left(\frac{{R}^2-d^2-y^2}{2 d y}\right)}$ if ${\min\(R_L,|R-d|\)} < y \leq {\min\(R_L,R+d\)} $ and $c_3(y,R)= \frac{\lambda_1\omega{P_1 } d^\alpha y^{-\alpha+1}}{P_2+{P_1 } d^\alpha y^{-\alpha}}$ otherwise, $u = \arccos\(\frac{d^2 + 2L R - {L}^2}{2 d R}\)$, $\mu=\min(\nu_\omega,u)$, and $\nu_\omega=\arccos \left( \frac{R\sin^2(\omega)-|\cos(\omega)|\sqrt{d^2-R^2\sin^2(\omega)}}{d}\right)$.				 
			\vskip 0pt \noindent
			{\bf Proof.} 
See Appendix G.
\hfill $\blacksquare$}\label{prop_OP3}
	\end{proposition}
\vspace {-5px}
Recall that the second and first parts of the OP \eqref{Eq:CondSIR3} imply the effects of the nearest interferer and the aggregate interference except the nearest one, respectively. The following remark specifies the effect of directional transmissions on OP.
\vspace {-10px}
\begin{remark}[Effect of Directional Transmissions on OP] 
	\emph{As the directional transmission beamwidth $\omega$ becomes smaller, the common interfering probability  \eqref{Eq:common_interfering_prob} and the thinning probability $\frac{\omega}{2\pi}$ decrease. This makes both of the first and second parts of the OP \eqref{Eq:CondSIR3} become one, namely, interference-free networks. This means narrow beamwidth decreases the spatial correlation of secondary users (i.e., the measured interference $I$ has less impact on the OP). On the other hand, as~$\omega$ becomes larger, the OP \eqref{Eq:CondSIR3} converges to that with blockage effect only (see Proposition~\ref{prop_OP2}).}
\end{remark} \vspace {-10px}
We approximate $\mathsf{P}$ by deriving the expected value of $R$: $\mathsf{P} \approx\mathcal{P}({R}_I, L, d, \beta, \omega)$. The value $R_I$ is calculated from \eqref{Eq:EmptyBallRadius_1} by thinning the primary TX density $\lambda_1$ with probability $\frac{\omega}{2\pi}$.\vspace{-5px}
{\begin{align}
\(\frac{I}{P_1} + \frac{\omega\lambda_1}{\alpha-2} {R_L}^{2-\alpha} \) {{R_I}}^\alpha-\frac{\omega\lambda_1}{\alpha-2}{{R_I}}^2-1=0.
\end{align}}
\vspace {-10px}
\vspace {-20px}
\subsection{Area Spectral Efficiency Maximization}
In this section, we seek to provide ASE-maximizing $\SIR$ target $\beta$ when blockage and directional transmission effects are considered. The ASE is given, via some modifications of the ASE without blockage, in Section~\ref{ase_wo_bl}, by 
considering the void probability that the nearest interferer is outside the range of $R_L$ as in \cite{jm_access}. Then, the ASE is represented as follows:\vspace {-10px}
\begin{align}
\mathcal{A} =  \exp\(-1\)\ln(1+\beta)
c^*\lambda_2\int_{[R-d]^+}^{R_L} & \mathcal{P}^2(r,L,d,\beta,\omega)  \frac{g_{R}(r)}{{1-\exp\(-\omega\lambda_1 {R_L}^2 /2\)}} dr,\label{ProblemReformulation_blockage}\end{align}\normalsize
where $g_{R}(r)=\omega\lambda_1 r e^{-\omega\lambda_1 r^2 /2}$.  As in \eqref{ASE_1_re}, we reformulate the optimization problem as:
\begin{align}
\underset{\beta\geq{\beta_{\min}}}{ \max}\ln(1+\beta) \beta^{-\frac{2}{\alpha}}\exp\[-\int_{[\tilde{s}-d]^+}^{R_L}c_3(y,\tilde{s}) dy\]^{\beta^{\frac{2}{\alpha}}},
\label{ProblemReformulation_blockage_re}\end{align}
where $\tilde{s}$ is a positive constant which satisfies $\mathcal{{P}}(\tilde{s},L,d,\beta,\omega) =\frac{\int_0^\infty \mathcal{{P}}^2(r,L,d,\beta,\omega)  g_{R}(r) dr}{ \int_0^\infty \mathcal{{P}}(r,L,d,\beta,\omega)  g_{R}(r) dr}$.
The minimum decoding $\SIR$ target $\hat{\beta}_{\text{min}}$ is derived by: 
\begin{align}
\hat{\beta}_{\text{min}}\!=\!\(\frac{\pi d^2 \lambda_1 {P_2}^{\frac{2}{\alpha}} \[\tau\!+\! \rho(\gamma,R_L)\tau-\rho(\gamma,R_L)\]}{ {P_1}^{\frac{2}{\alpha}} \gamma^{\frac{2}{\alpha}} (1-\tau) }\)^{\frac{\alpha}{2}},
\end{align}
representing that $\hat{\beta}_{\text{min}}$ decreases with the blockage density and size.
In the above-$6$ GHz case, with the same technique used at Proposition~\ref{prop_optimal_beta}, the corresponding optimal decoding target $\beta^*$ can be also represented as $\max(\beta_1,\hat{\beta}_{\text{min}}), $ when $\beta_1$ is a value that satisfies $-2\int_{[\tilde{s}-d]^+}^{R_L} {c_3(y,\tilde{s})} dy {\beta_1}^{\frac{2}{\alpha}} \ln(1+\beta_1) + \frac{\alpha\beta_1}{1+\beta_1} - 2\ln(1+\beta_1)$.
\vspace {-10px}
\begin{remark}[Effect of Blockages] \emph{
The optimal density of concurrent secondary transmissions, $\Lambda_2^*=\frac{1}{\pi d^2 \rho_0(\beta^*,R_L) }$, depends on blockage density and size. Large or densely deployed blockages reduce the LOS distance $R_L$, which in turn reduces $\rho_0(\beta^*,R_L)$. Also, the optimal density $\Lambda_2^*$ increases; thus, the optimal decoding target $\beta^*$ should be decreased to retain the optimal transmission density. 
}
\end{remark}
\vspace {-10px}
\vspace {-10px}
\begin{remark}[Effect of Beamwidth] \emph{As the beamwidth $\omega$ goes to $0$, the optimal decoding target becomes $ \frac{\alpha\beta^*}{1+\beta^*} - 2\ln(1+\beta^*)=0$. This implies that when the beamwidth is extremely small, the optimal target $\beta^*$ is no longer affected by the primary interference.}
\end{remark} 
\vspace {-10px}

\vspace {-10px}
\section{Performance Evaluations} \label{Performance Evaluations}

In this section, the analytical SaP design results are evaluated using USRP testbed and MATLAB simulations.
\vspace {-15px}
\subsection{Opportunistic Probability Verification by USRP Testbed Experiments}
\begin{figure*}   
\centering
  \subfigure[The USRP testbed setup for CR networks]{\centering 
   \includegraphics[width=7.3cm]{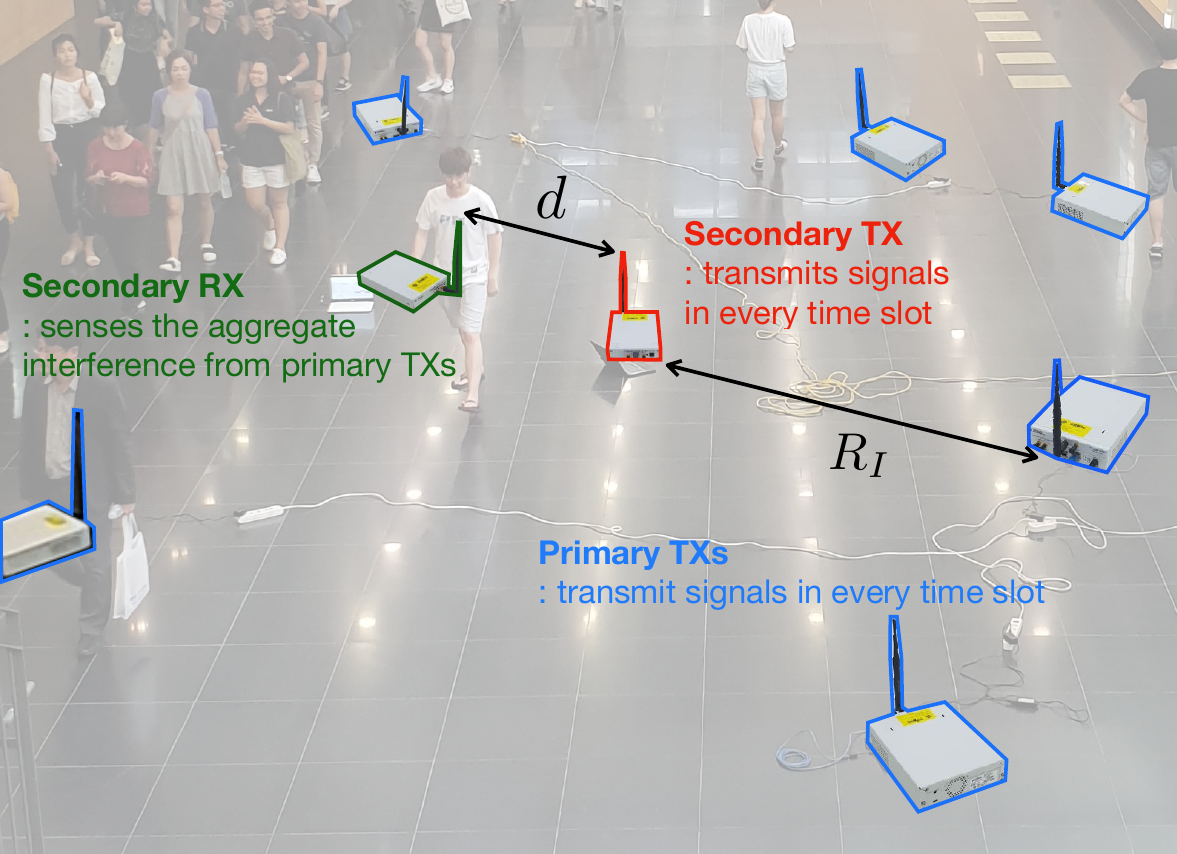} }
   \qquad
  \subfigure[OP comparison]{
   \includegraphics[width=7.5cm]{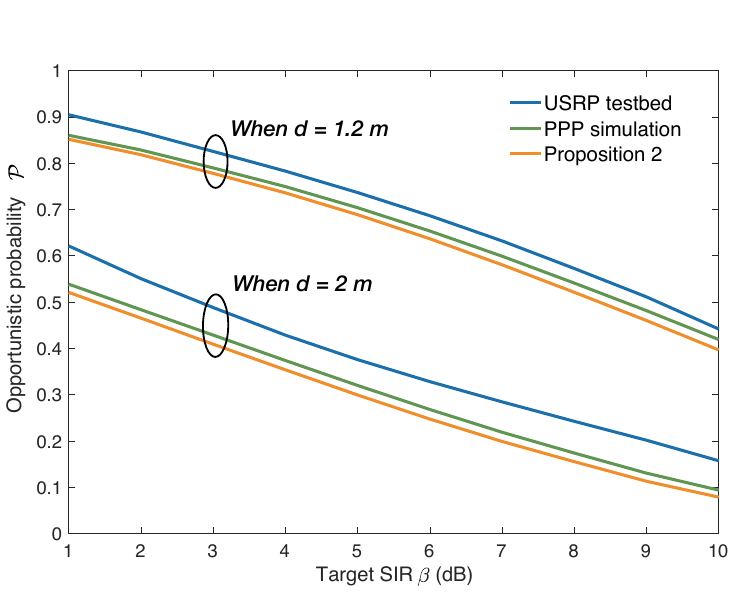}} \vskip-10pt
\caption{OP with SaP according to target $\SIR$ $\beta$ ($R_I = 3.6 \text{ m}$, $\lambda_1= 7 \times 10^3 \text{ TXs}/\text{km}^2$, $P_1=6.06$ dBm, $P_2=3.162$ dBm, $\alpha=3$). \label{fig:CondSIR}} \vspace{-.3in}
\end{figure*}

To verify the derived OP in Proposition~\ref{prop_OP1}, we compare the values with measurements obtained using a real testbed featuring eight pairs of laptops and USRPs (two {NI-USRP $2900$}, six {NI-USRP $2922$}), as shown in Fig.~\ref{fig:CondSIR}(a). Each laptop is connected to NI-USRP $2900$ via USB $3.0$ connection and NI-USRP $2922$ via 1-Gbps Ethernet cable. Each laptop utilizes LabVIEW Communications $2.0$ software. We investigate four kinds of primary network topologies with different deployments comprising six USRPs. Every USRP representing the primary network always turns on and transmits signals with constant power $P_1=6.06$ dBm at a center frequency of $2.4$ GHz in every time slot with the bandwidth of $200$ kHz, which is a realistic setting in the conventional cellular network. In the meantime, the secondary network comprises two USRP as a pair of secondary TX and RX. Secondary TX transmits a signal of $P_2=3.162$ dBm at a center frequency of $2.4$ GHz in every time slot, with a bandwidth of $200$ kHz. The distances between secondary TX and secondary RX are $1.2$m, and $2$m. The laptops are used to configure the paired USRP devices. The distance $R$ between the secondary TX and the nearest primary TX is $3.6$ m. The secondary RX measures the transmitted signal from the secondary TX and the aggregate interference from the primary TXs with a sensing bandwidth of $600$ kHz, and passes the signal to a lowpass filter with a bandwidth of $400$ kHz. The RX gain is $52$ dB. The value of $\SIR$ is obtained by calculating the measured power difference between the signal and the aggregate interference.

We also performed MATLAB simulations. We considered a square of $1\times1$ $\text{km}^2$ where the primary TX density was $\lambda_1$, the transmission powers $P_1$ and $P_2$, and the distance $R_I$ identical to that of the USRP testbed experiments. The path-loss exponent was set to $3$ considering that the testbed experiments are performed inside the building (indoor).
Fig.~\ref{fig:CondSIR}(b) shows that the analytical OP from Proposition~\ref{prop_OP1} is relatively consistent with that of the PPP simulation.
A slight difference is apparent, but the tendencies are similar. 
On the other hand, it is observed that the measured OP is higher than the analysis counterpart. 
The reason is that two primary TXs do not adequately represent a real primary network comprising a near-infinite number of TXs. This renders secondary RX measures of aggregate interference lower than those of analysis and PPP simulation, yielding an increment in $\SIR$ coverage. This gap would be mitigated by increasing USRP numbers and/or the experimental area.

\vspace{-10px}
\subsection{Opportunistic Probability and Area Spectral Efficiency Maximization in below-$6$ GHz}
This section verifies the OP and ASE calculations in the absence of blockage and beamforming (i.e., in the below-$6$ GHz scenario). The simulation parameters had the following default settings unless specified otherwise: Transmission powers of the primary and secondary TXs $P_1=43$ dBm and $P_2=23$ dBm, respectively, and a path-loss exponent $\alpha=4$.


\begin{figure*}
\centering
  \subfigure[OP according to the measured interference $I$ ($\lambda_1= 5 \times 10^2 \text{ TXs}/\text{km}^2$, $d = 2$ m).]{\centering
   \includegraphics[width=7.5cm]{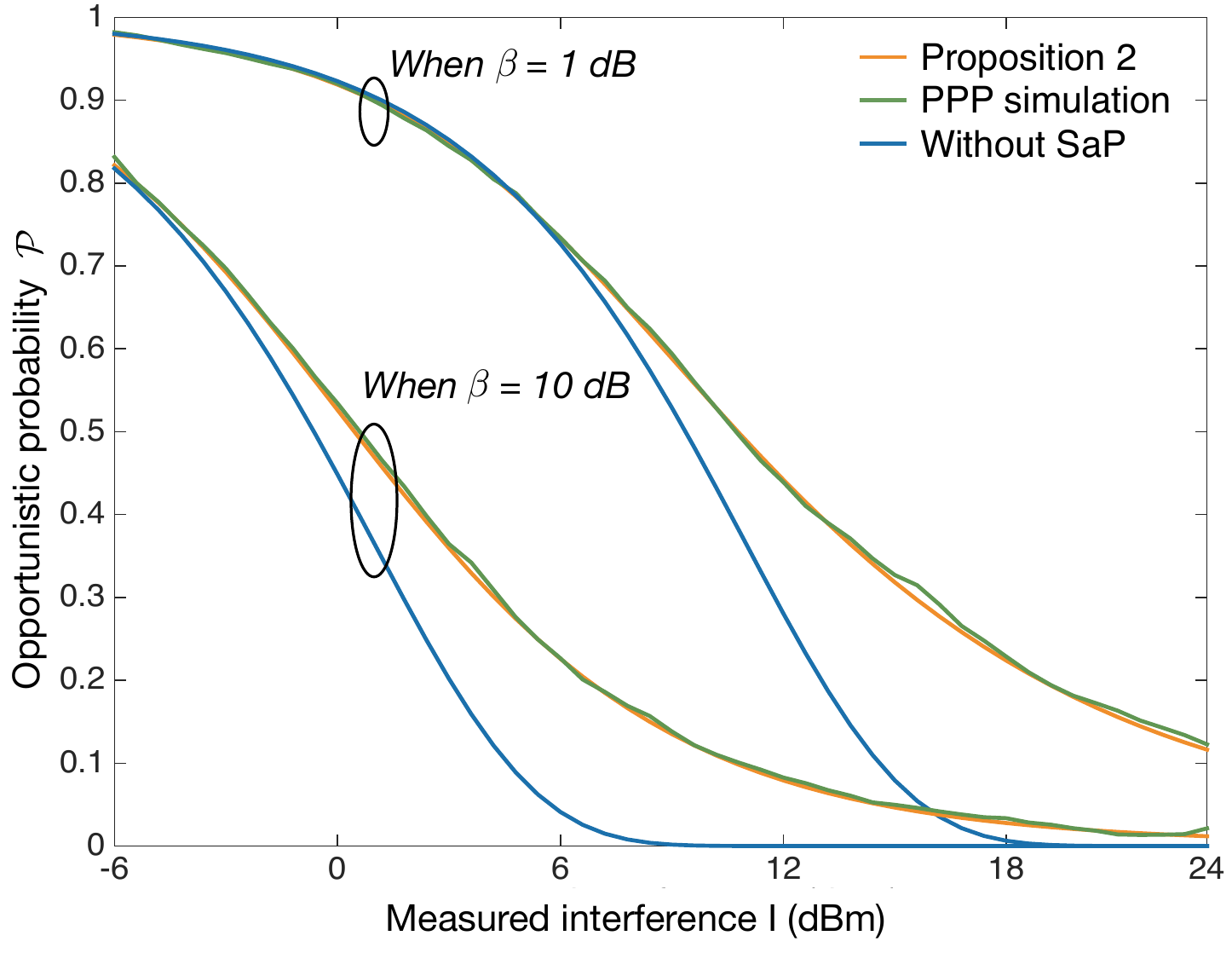} \label{fig:CondSIR_ri}}
   \quad
  \subfigure[Numerically evaluated ASE ($\lambda_1= 80 \text{ TXs}/\text{km}^2$, $\lambda_2= 1.6 \times 10^4 \text{ TXs}/\text{km}^2$, $d = 3$ m).]{
   \includegraphics[width=7.3cm]{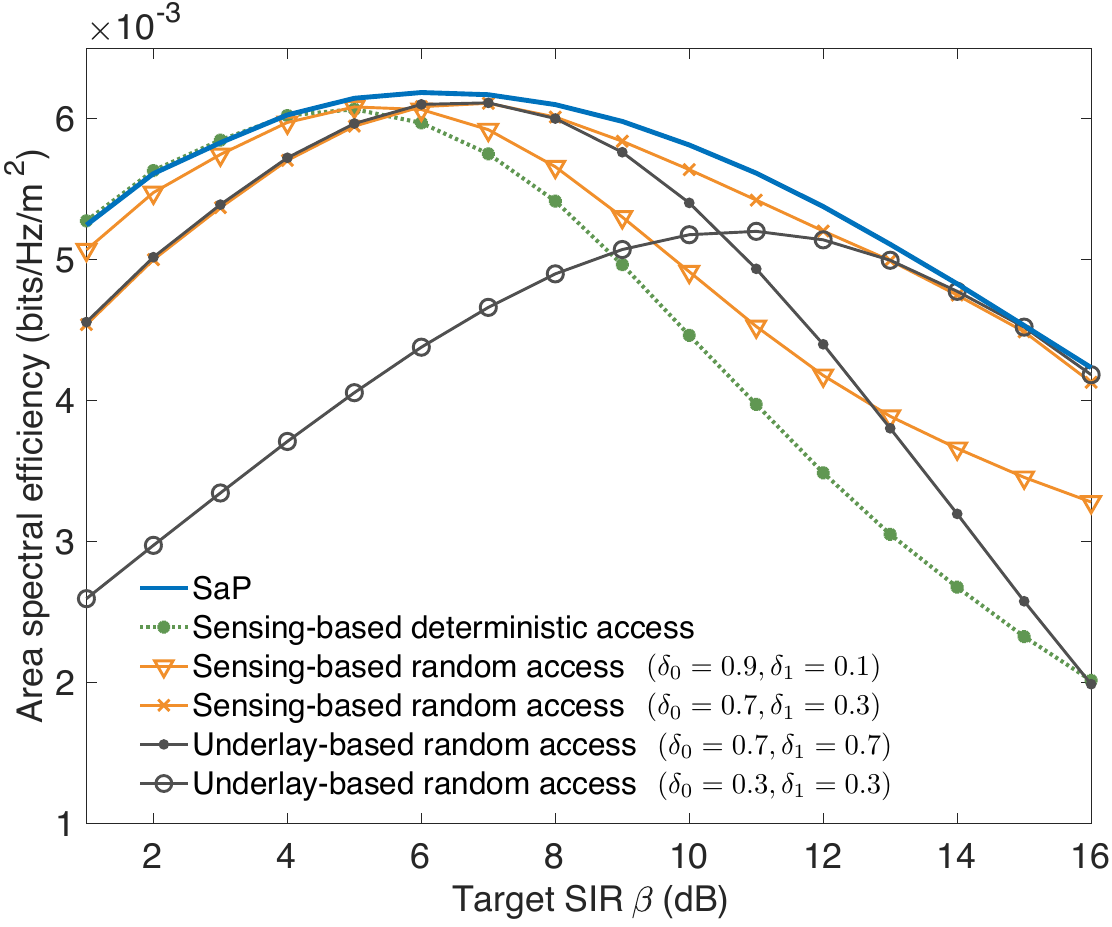}\label{fig:ase_analysis_ppp}}\vskip-10pt
\caption{OP and ASE of SaP in below-$6$ GHz Spectrum. In (b), in-band underlay random access \cite{d2d_2} and sensing-based random access \cite{d2d} are represented. Each scheme is evaluated with two different cases.} 
\vspace{-.3in}
\end{figure*}




Fig.~\ref{fig:CondSIR_ri} presents the OP in terms of the measured interference $I$. 
As $I$ increases, the OP decreases, showing that the analytical results were in good agreement with the simulated results. 
Here, the OP in the absence of SaP represents the predicted $\SIR$ coverage when the interference level at RX was identical with that at TX, $I$. 
The OP gap with and without SaP increases with $I$ because of the weakened interference correlation between the secondary TX and the RX (see Corollary~\ref{asmyptotic_OP}). This implies that SaP can reduce the number of false alarms (when the secondary TX decides not to access the spectrum because of measured high interference, even if the interference is in fact low).
Fig.~\ref{fig:ase_analysis_ppp} compares the ASE of SaP with the state-of-art MAC protocols, underlay-based random access \cite{d2d_2}, spectrum sensing-based random access \cite{d2d}, and spectrum sensing-based deterministic access, each of which is described as follows. 
\begin{enumerate}
\item \textbf{Underlay-based random access \cite{d2d_2}:} A transmitter accesses the medium with probability~$\delta$. Otherwise, it remains silent. 
\item \textbf{Spectrum sensing-based random access \cite{d2d}:} A transmitter accesses the medium with probability $\delta_0$ when the measured level of interference is less than a given threshold. Otherwise, it can access the medium with probability $\delta_1$, where $\delta_0\geq \delta_1$ without loss of generality. 
\item \textbf{Spectrum sensing-based deterministic access:} A transmitter accesses the medium only when the measured level of interference is less than a given threshold. 
\end{enumerate}
The thresholds of the spectrum sensing schemes are fine-tuned for fair comparisons. It is shown that SaP always outperforms the benchmarks, confirming its near-to-optimal performance by the well-adjusted one-to-one mapping function explained in Sec.~\ref{subsection:optimality}, which provides a graceful tradeoff between the access probability and the measured interference level.

\vspace{-.2in}
\subsection{Opportunistic Probability and Area Spectral Efficiency Maximization in above-$6$ GHz}\label{sec:ase_blockage_simul}

\begin{figure*}   
\centering
  \subfigure[OP through simulation when $\xi=0.04$]{\centering 
   \includegraphics[width=7.5cm]{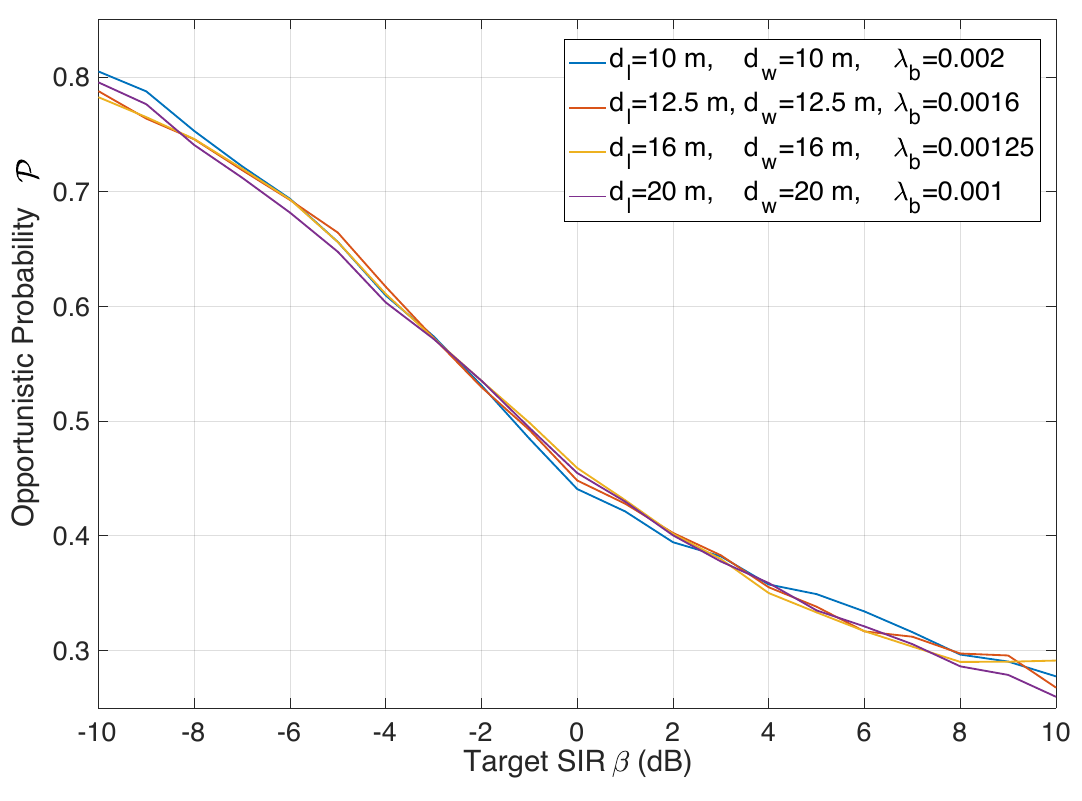} \label{fig:pa_veri}}
   \quad
  \subfigure[Optimal $L$ according to $\xi$ and its fitted curve]{
   \includegraphics[width=7.5cm]{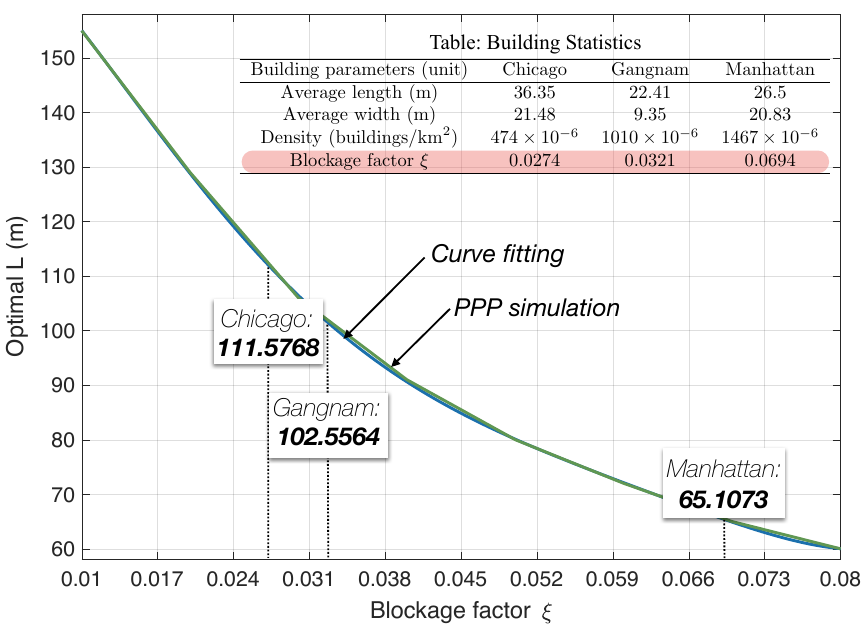}\label{fig:opt_da_fac}}\vskip-10pt
\caption{OP in above-$6$ GHz and optimal distance of joint unblocked ellipse area, $L$ ($\lambda_1= 80 \text{ TXs}/\text{km}^2$, $\lambda_2= 1.6 \times 10^4 \text{ TXs}/\text{km}^2$, $d = 5$ m)}\vspace{-.3in}
\end{figure*}

The simulation parameters are identical except a path-loss exponent of $\alpha=2.7$, which is changed to reflect the effects of mmW.
Before comparing analytical and simulation results, it is necessary to determine the axis length $L$ in \eqref{Eq:CondSIR2} using 2D blockage geometrical information. Recalling that LOS condition depends on $\lambda_b\(d_l+d_w\)$ (see \eqref{Eq:JointLOSProb}), we first define a blockage factor $\xi=\lambda_b\(d_l+d_w\)$. We then check whether it is appropriate to use the blockage factor $\xi$ in simulations with different values of $\lambda_b$, $d_l$, and $d_w$ that yield the same $\xi$ (see Fig.~\ref{fig:pa_veri}). We then find the axis length $L$ that affords the smallest gap between the OP results from PPP simulation and analysis of \eqref{Eq:CondSIR2}, and create a fitted curve of optimal $L$ against $\xi$ as follows: $L = 2.051\times10^8 \xi^5 -4.729\times10^7\xi^4+3.847\times10^6\xi^3-1.118\times10^5\xi^2-1318\xi+176 $, where the \emph{sum of squared errors} (SSE) is $0.4103$. Armed with this result, we calculate the optimal $L$s of three cities: Chicago, Gangnam, and Manhattan, as shown in Fig.~\ref{fig:opt_da_fac}. Relevant parameters (density, and average length and width) are summarized in the Table of Fig.~\ref{fig:opt_da_fac}.

\begin{figure*}   
\centering
  \subfigure[SaP OP]{\centering 
   \includegraphics[width=7.5cm]{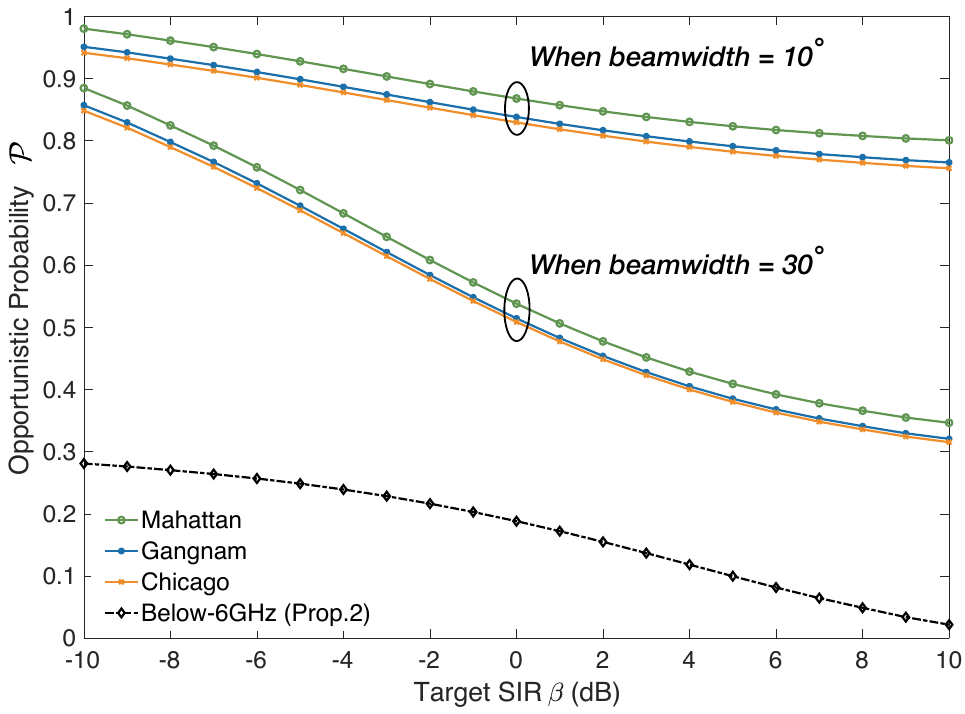} \label{fig:blockage_ap}}
   \quad
  \subfigure[ASE of Manhattan scenario when $\omega=10^{\circ}$]{
   \includegraphics[width=7.5cm]{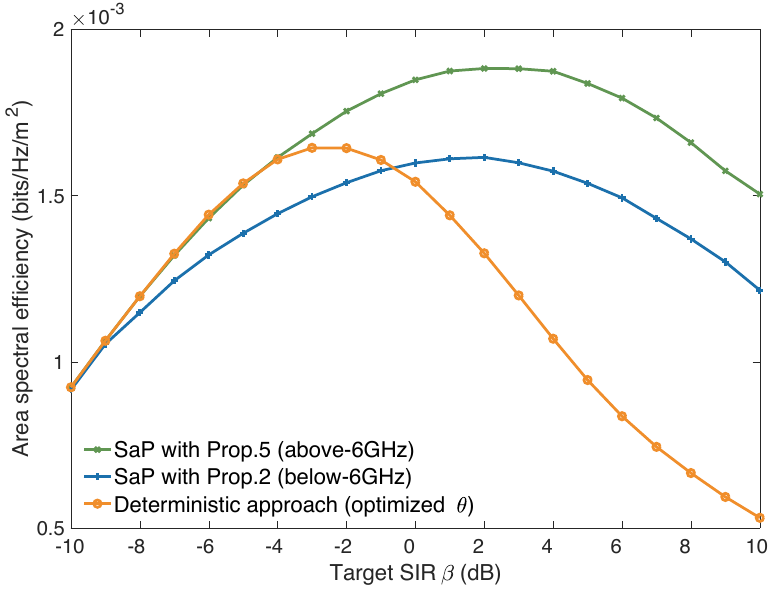}\label{fig:blockage_ase}}\vskip-10pt
\caption{OP and ASE of SaP ($\lambda_1= 80 \text{ TXs}/\text{km}^2$, $\lambda_2= 1.6 \times 10^4 \text{ TXs}/\text{km}^2$, $d = 5$ m)}\vspace{-.3in}
\end{figure*}

Fig.~\ref{fig:blockage_ap} illustrates OP according to the three city scenarios with two different beamwidth, i.e., $\omega=\frac{\pi}{18}$ and $\omega=\frac{\pi}{6}$. OP of three cities varies because of the blockage factor difference as shown in Fig.~\ref{fig:opt_da_fac}.
As the decoding target increases, the overall OP decrease. This makes the secondary TXs discourage themselves to access the channel, so as to maximize ASE of the secondary network. 
It also shows that the OP of Proposition~\ref{prop_OP1} is quite different from those of Proposition~\ref{prop_OP2} and the simulation, emphasizing the blockage effect. The OPs of three cities increase as the beamwidth of the primary network decreases, yielding the directional transmission effect.
Fig.~\ref{fig:blockage_ase} verifies the ASE improvement with the SaP algorithm, as revealed by MATLAB simulation.
In addition, using the OP from Proposition~\ref{prop_OP1} yields lower ASE, implying that OP predicted without considering the blockages decreases the secondary performance than when using the TX sensing result directly.
This indicates that considering the blockage and beamforming effects is highly important to provide an appropriate OP, especially in an urban area like Manhattan.

\vspace{-.25in}
\section{Conclusion}
\vspace{-5px}
We tackled the problem of detecting spectrum access opportunity in CR networks posed by the interference gap between the secondary TX and the RX, caused by differences in location and exposed nodes. 
We developed an SaP protocol whereby the interference level at the secondary RX is predicted based on the channel measurement by its paired TX. 
SG has been applied to quantify the spatial interference difference in the form of the $\SIR$ coverage probability, defined as the OP, which is directly used for ALOHA-based random access with optimization. 
Based on the SaP framework, the relations between access threshold and target $\SIR$ are identified with the mapping function to maximize secondary network ASE while protecting the service quality of the primary network, providing useful insights in terms of SaP design. 
The proposed SaP framework provides a useful guideline to realize massive connectivity of 5G such that given the local interference information, each device enables to control its access decision optimally to maximize the spectral efficiency without the help from a centralized controller. 
This work can be extended in several interesting ways to further improve the SaP framework. First, we could consider cooperative sensing among secondary TXs, which is known to outperform non-cooperative sensing, but at a cost of increased energy consumption. 
Second, the use of machine-learning techniques to track environmental changes would align with the recent research interests of those studying wireless communications. 
Last, the use of more advanced physical layering techniques (e.g., compressive sensing and full duplexing) is promising.
\vspace{-.2in}
\section*{Appendix}

\vspace{-.1in}
\subsection{Proof of Proposition~\ref{prop_OP1}} \label{prop_OP_proof}
\vspace{-5px}
Consider a typical secondary TX and its paired RX located at a distance of $d$. Let $x$ denote the distance between the RX and the secondary TX's nearest primary TX $T_1$ (see Fig.~\ref{deployment}). Then the OP is represented as:\vspace{-5px}
\begin{align}
&\Pr\[\frac{P_2 h^{(0)} d^{-\alpha}}{P_1 \(h x^{-\alpha} + \sum_{i\in \Phi_1 \setminus T_1} h^{(i)} {x_1 ^{(i)}}^{-\alpha}\)} \geq \beta\] \\ 
&= \mathsf{E}_{x,h} \[\exp\(-\frac{\beta P_1d^\alpha h x^{-\alpha}}{P_2} \)\] \times \mathsf{E}_{\Phi_1, \{h^{(i)}\}} \[\exp\(- \frac{\beta P_1 d^\alpha \sum_{i\in \Phi_1 \setminus T_1} h^{(i)} {x_1 ^{(i)}}^{-\alpha}}{P_2} \)\]. \label{prop1_def}
\end{align}
Utilizing the fading $h \sim \exp(1)$ provides \vspace{-5px}
\begin{align}
 \mathsf{E}_{x,h} \[\exp\(-\frac{\beta P_1h d^\alpha x^{-\alpha}}{P_2} \)\] &=\mathsf{E}_x\[ \frac{P_2}{P_2 + {\beta P_1d^\alpha x^{-\alpha}}}\]  \\ &\stackrel{(a)}=\mathsf{E}_\nu\[ \frac{P_2}{P_2 + {\beta P_1d^\alpha \({R}^2 - 2 d {R} \cos(\nu) + d^2\)^{-\frac{\alpha}{2}}}}\] \label{nu} \\&\stackrel{(b)}=\frac{1}{2\pi} \int_0^{2\pi} \frac{P_2}{P_2 + P_1 \beta d^\alpha \({R}^2 - 2 d {R} \cos(\nu) + d^2\)^{-\frac{\alpha}{2}}} d\nu, \label{ti}
\end{align}
\noindent 
where (a) follows from the triangular relation $x^2={R}^2 - 2 d {R} \cos(\nu) + d^2$ with $\nu$ representing the angle between $T_1$ and a typical secondary RX (see Fig.~\ref{deployment}), and (b) follows from the fact that $\nu$ is uniformly distributed over 2$\pi$.
Under the empty ball condition, there is an empty ball of radius $R$ with no primary TX inside. Imagine a thin circular ring $B_y$ of radius $y$ around the typical secondary RX. The secondary RX does not have any interferer within the area intersecting $B_y$ and the empty ball. Then, the intensity function of the interferer becomes:
\begin{figure}
	\centering 
	{\includegraphics[width=7cm]{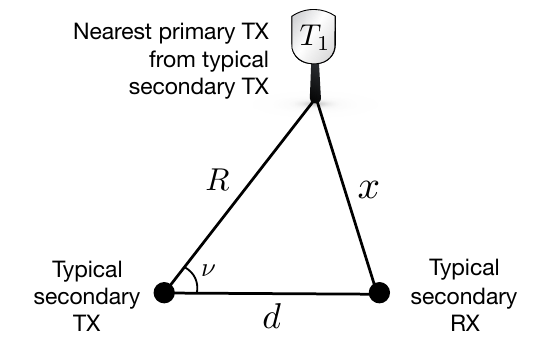}} \vskip-10pt
	\caption{A network topology comprising a pair of typical secondary TX and RX with the nearest primary TX.\label{deployment}}\vspace{-.3in}
\end{figure}
\vspace{-.1in}
\begin{align}
\lambda'= \begin{cases} 
0 & {\textrm{if $ 0< y \leq \max(0,R -d)$,}} \\
2 \arccos\(\frac{{R}^2-d^2-y^2}{2 d y}\) \lambda_1 y&{\textrm{if $\max(0,{R}-d) < y \leq {R}+d$, }} \\
2 \pi \lambda_1 y &{\textrm{if $R+d < y$.}}
\end{cases}\label{int_func}
\end{align} 
Applying \eqref{int_func} and probability generating function of PPP, the right part of \eqref{prop1_def} becomes\vspace{-5px}
\begin{align}\nn
\mathsf{E}_{\Phi_1, \{h^{(i)}\}} &\[\exp(- \frac{\beta P_1 d^\alpha \sum_{i\in \Phi_1 \setminus T_1} h^{(i)} {x_1 ^{(i)}}^{-\alpha}}{P_2} )\] \\
\nn
=&\exp\left(-\lambda_1\int_{R+d}^{\infty}\frac{2\pi P_1 \beta d^\alpha y^{-\alpha+1}}{P_2+P_1 \beta d^\alpha y^{-\alpha}}dy - \lambda_1\int_{|{R}-d|}^{{R}+d} \frac{2 \arccos\left(\frac{{R}^2-d^2-y^2}{2 d y}\right)P_1 \beta d^\alpha y^{-\alpha+1}}{P_2 + P_1 \beta d^{\alpha} y^{-\alpha}} dy\) \\ 
&\ \times \exp\(-\lambda_1 \int_{\max(0,{R}-d)}^{|{R}-d|}\frac{2\pi P_1 \beta d^\alpha y^{-\alpha+1}}{P_2 +P_1 \beta d^\alpha y^{-\alpha}}dy\), \label{condsir_right}
\end{align} 
where $y$ represents the distance between a typical secondary RX and primary TXs. 
Since $T_1$ is the nearest primary TX only from the perspective of the secondary TX, there is a probability that a primary TX is located nearer than $T_1$ from the perspective of the secondary RX.
Thus we integrate the distance $y$ from $0$ to $\infty$.
Plugging \eqref{ti} and \eqref{condsir_right} into \eqref{prop1_def} and applying mapping theorem (i.e., an aggregate interference from a PPP set with density $\lambda$ and power $\beta P$ is the same with that from a PPP set with density $\beta^{\frac{2}{\alpha}}\lambda$ and power $P$, almost surely \cite{Haenggi13}) finalize the proof.

\vspace{-.2in}
\subsection{Proof of $R_I$ without Blockage Effect}\label{r_i_proof}
\vspace{-5px}
When the distance between a typical secondary TX and its nearest primary TX is equal to $R_I$, the expected interference is represented using Campbell's theorem \cite{Haenggi13} as:\vspace{-5px}
\begin{align}
{P_1\mathsf{E}_h\[h {R_I}^{-\alpha}\]}+ \underbrace{P_1 \mathsf{E}_{\Phi_1, \{h^{(i)}\}}\[\sum_{i \in \Phi_1 \setminus T_1} h^{(i)} {x_1^{(i)}}^{-\alpha}\]} _{(a)}, \label{avg_intf}
\end{align}
where $T_1$ denotes the nearest primary TX.
Let $I_r$ denote $\sum_{i \in \Phi_1 \setminus T_1} h^{(i)} {x_1^{(i)}}^{-\alpha}$ in $(a)$.
In order to derive $(a)$, we consider the Laplace transform of $I_r$, $\mathcal{L}(s) = \mathsf{E}\[e^{-s I_r}\]$.\vspace{-5px}
{\begin{align} \label{R_I_prop1}
\mathcal{L}(s) \nn&= \mathsf{E}_{\Phi_1} \[\prod_{i\in \Phi_1 \setminus T_1} \mathsf{E}_{h^{(i)}} \(e^{-s h^{(i)} {x_1^{(i)}}^{-\alpha}}\)\] \overset{(b)}{=} \exp\(-2\pi\lambda_1\mathsf{E}_h\[\int_{R_I}^\infty \(1-\exp\(-s h r ^{-\alpha}\)\) r dr\]\)\\ 
&\overset{(c)}{=}\exp\(-\pi\lambda_1\mathsf{E}_h\[\int_0^{{R_I}^{-\alpha}}\frac{ t^{-\frac{2}{\alpha}} s h dt}{ e^{s h t}} - r^2 \(1-e^{-s h r^{-\alpha}}\)\]\) 
\end{align}}\noindent
\noindent where $(b)$ follows from the probability generating function of PPP \cite{Haenggi13} and $(c)$ from the partial integral.
By applying that $ \mathsf{E}\[I_r\] = -\frac{\partial}{\partial s}\mathcal{L}(s) |_{s=0}$, the function $(a)$ is represented as 
$
\frac{2 P_1 \pi\lambda_1 \mathsf{E}\[h\] {R_I}^{2-\alpha}}{\alpha-2}.
$
Now that $h \sim \exp(1)$, finding $R_I$ that satisfies $I = P_1 {R_I}^{-\alpha} +\frac{2 P_1 \pi\lambda_1 {R_I}^{2-\alpha}}{\alpha-2}$ finalizes the proof.
\vspace{-.2in}
\subsection{Proof of Lemma~\ref{lemma_avgOP}}
\vspace{-5px}
Now that the signal experiences Rayleigh fading, the success probability is divided into:\vspace{-5px}
\begin{align}
\Pr\(\SIR_2 \geq \beta |I \) =\Pr\(\frac{h^{(0)} d^{-\alpha} P_2}{ \sum I_1} > \beta |I \) \Pr\(\frac{h^{(0)} d^{-\alpha} P_2}{\sum I_2} > \beta |I \),
\end{align}
where $\sum I_1$ denotes the sum of interference from primary TXs and $\sum I_2$ the sum of interference from secondary TXs, respectively.
The first probability is directly calculated from the OP.
For deriving the second probability, we assume that transmitting secondary TXs are independently thinned with $\overline{\mathcal{P}}(d,\beta)=\frac{1}{\pi \lambda_2 d^2 \rho_0(\beta,\infty)}$.
Applying a thinned secondary concurrent transmitting density $\frac{1}{\pi d^2 \rho_0(\beta,\infty)}$ in $\SIR$ coverage \cite{Andrews11} completes the proof.
 
 \vspace{-.2in}
\subsection{Proof of Lemma~\ref{lemma_mintheta}}
From the outage probability derived by using the Theorem 2 in \cite{Andrews11}, we represent the outage probability of the primary network as:\vspace{-5px}
\begin{align}
\Pr\[\SIR_1< \gamma\] &= 1-\frac{\lambda_1{P_2}^{\frac{2}{\alpha}}}{\lambda_2 \overline{\mathcal{P}}(d,\beta)\rho_0(\gamma,\infty) {P_1}^{\frac{2}{\alpha}}+\lambda_1 {P_2}^{\frac{2}{\alpha}}( \rho(\gamma,\infty) + 1) } \\
&= 1-\frac{\lambda_1{P_2}^{\frac{2}{\alpha}}}{  \frac{\rho_0(\gamma,\infty) {P_1}^{\frac{2}{\alpha}}}{\pi d^2 \rho_0(\beta,\infty)}+\lambda_1 {P_2}^{\frac{2}{\alpha}}( \rho(\gamma,\infty) + 1) }, \label{outage_final}
\end{align}
where $\rho_0(x,t):=x^{\frac{2}{\alpha}}\int_0^{t}\frac{du}{1+u^\frac{\alpha}{2}}$ and $\rho(x,t):=x^{\frac{2}{\alpha}}\int_{x^{-\frac{2}{\alpha}}}^{t}\frac{du}{1+u^\frac{\alpha}{2}}$.
Reminding the protection condition that $\Pr\[\SIR_1< \gamma\] \leq \tau$, the maximum threshold $\beta$ is given by finding $\beta$ which makes \eqref{outage_final}$=\tau$.

 \vspace{-.2in}
\subsection{Proof of Proposition~\ref{prop_optimal_beta}}
Note that only two solutions satisfy the first derivative of ASE when the path-loss exponent is larger than two, i.e., $\alpha>2$.  One is zero and the other is $\beta^*$, which is always larger than zero, thus it becomes optimal decoding target when $\beta^*>\beta_{\min}$. That is because the second derivative of ASE has an unique solution when $\beta$ is smaller than $\beta^*$, and it is bounded by negative zero as $\beta$ goes to infinity, i.e., there is no more inflection point. This means the result in Proposition~\ref{prop_optimal_beta} is always valid under the minimum decoding target and path-loss condition.

\vspace{-.2in}
\subsection{Proof of Proposition~\ref{prop_OP2}}
\vspace{-5px}
Let $T_1$ denote the nearest primary TX. By reference to the empty ball radius $R$, we consider three cases.
First, when $R<\frac{L}{2}-\frac{d}{2}$, $T_1$ is in the joint unblocked region. Thus the signal from $T_1$ interferes with the secondary RX, the interference power of which is derived using a triangular function as in \eqref{ti}. Next, when $\frac{L}{2}-\frac{d}{2}\leq R<\frac{L}{2}+\frac{d}{2}$, $T_1$ interferes with the secondary RX if that RX is within the ellipse's area (the purple region in Fig.~\ref{fig:blockage_t1}). Otherwise, $T_1$ does not interfere with the secondary RX, causing an exposed node problem (red region in Fig.~\ref{fig:blockage_t1}). The angle $u$ in Fig.~\ref{fig:blockage_exp} is derived from $d^2-2 d R \cos(u) + {R}^2 = (L-R)^2$ as follows:\vspace{-5px}
\begin{align}
u =\arccos\(\frac{d^2 + 2L R - {L}^2}{2 d R}\) \label{angle_x}
\end{align}
Thus, when the angle $\nu$ in \eqref{ti} is smaller than $u$ of \eqref{angle_x}, the primary TX interferes with the secondary RX with an interference power derived as in \eqref{ti}.
Lastly, when $\frac{L}{2}+\frac{d}{2}\leq R$, $T_1$ cannot interfere with the secondary RX. Thus, the interference power becomes $0$. In terms of aggregate interference from primary TXs outside the empty ball, the fact that interferers located farther than $R_L$ from the secondary RX cannot interfere yields the final result.

\vspace{-.2in}
\subsection{Proof of Proposition~\ref{prop_OP3}}
\vspace{-5px}
In the same manner as the proof of Proposition~\ref{prop_OP2}, we consider three cases and apply the common interfering probability $p_c$ in \eqref{Eq:common_interfering_prob} to each case.

First, when $R<\frac{L}{2}-\frac{d}{2}$ (see case 1 in Fig.~\ref{fig:directional_exposed_2}), $T_1$ might cause exposed node problem because it is in the joint unblocked region. We hereafter let the value $\nu$ satisfying $\nu=\psi^{-1} (\omega)$ denote~$\nu_\omega$. When $\nu<\nu_\omega$, $T_1$ is not an interferer to the secondary RX. When $\nu \geq \nu_\omega$ on the other hand, $T_1$ becomes a common interferer to the secondary users with probability $p_c(\nu_\omega)$ and an exposed node with probability $1-p_c(\nu_\omega)$. Accordingly, the left part of \eqref{prop1_def} becomes \vspace{-5px}
\begin{align}
\frac{1}{\pi} \int_{0}^{\pi} \left((1-p_c(\nu))+\frac{ p_c(\nu) P_2 }{P_2 + P_1 \beta d^\alpha \({R}^2 - 2 d {R} \cos(\nu) + d^2\)^{-\frac{\alpha}{2}}}\right)d\nu.
\label{directional_ptx1}
\end{align}
The common interfering probability $p_c(\nu)$ is shifted at the boundary points such as $\nu=\nu_\omega$.
Second, when $\frac{L}{2}-\frac{d}{2}\leq R<\frac{L}{2}+\frac{d}{2}$, $T_1$ might cause exposed node problem depending on its angular correlation. Depending on $R$, there can be two angles corresponding to each boundary points, the larger one is denoted by $\nu_\omega^+$ and the smaller one is denoted by $\nu_\omega^-$ respectively (see Fig.~\ref{fig:directional_exposed_2}). We consider the blockage effect with $u$ in Fig.~\ref{fig:blockage_exp} and the directional transmission with $\nu_{\omega}^\pm$ at the same time. As a result, the left part of \eqref{prop1_def} becomes \vspace{-5px}
\begin{align}
\frac{\pi-\min(\nu_\omega,u)}{\pi}+\frac{1}{\pi} \int_{0}^{u} \left((1-p_c(\nu))+\frac{ p_c(\nu) P_2 }{P_2 + P_1 \beta d^\alpha \({R}^2 - 2 d {R} \cos(\nu) + d^2\)^{-\frac{\alpha}{2}}}\right)d\nu.
\label{directional_ptx2}
\end{align}
Lastly, when $\frac{L}{2}+\frac{d}{2}\leq R$, $T_1$ does not interfere with the secondary RX, always causing exposed node problem by the effect of the blockages.

Now that all the TXs transmit directional signals with beamwidth $\omega$, the intensity function of the rest of the interferers become $\lambda'= \frac{\lambda_1\omega}{2\pi}$ by the thinning probability $\frac{\omega}{2\pi}$. Applying this intensity $\lambda'$ and the probability generating function of PPP, the right part of \eqref{prop1_def} becomes \eqref{condsir_right}, where $\lambda_1$ is substituted by $\lambda'$. \vspace{-5px} 
Plugging \eqref{directional_ptx1},\eqref{directional_ptx2} and thinned \eqref{condsir_right} into \eqref{prop1_def} gives the desired result.

%
%

\end{document}